\documentclass[%
 apsrev4-2,prb,
 amsmath, amssymb, aps,
twocolumn
]{revtex4-2}

\usepackage{graphicx}
\usepackage{dcolumn}
\usepackage{bm}
\usepackage{amsmath}
\usepackage{graphicx, caption, subcaption}
\usepackage{xcolor}
\usepackage{soul}

\begin{document}

\title{Interplay of electric and magnetic fields in skyrmion phases \\
 of the classical Heisenberg model on a square lattice}

\author{A. Vela Wac$^{1,2,*}$}
\thanks{$^*$Corresponding author: avelawac@iflysib.unlp.edu.ar}
\author{F. A. Gómez Albarracín$^{1,2}$}%
\author{D. C. Cabra$^{1,3}$}

\affiliation{$^1$Instituto de Física de Líquidos y Sistemas Biológicos (IFLYSIB), UNLP-CONICET, Facultad de Ciencias Exactas, La Plata, Argentina,}
\affiliation{$^2$Departamento de Ciencias Básicas, Facultad de Ingeniería,
Universidad Nacional de La Plata, La Plata, Argentina}
\affiliation{$^3$Departamento de Física, Facultad de Ciencias Exactas,
Universidad Nacional de La Plata, La Plata, Argentina.}

\date{\today}
\makeatletter
\def\@fnsymbol#1{}
\makeatother

\begin{abstract}

Magnetic skyrmions are topologically stable spin textures that can be stabilized by Dzyaloshinskii–Moriya interaction and manipulated by external fields, making them promising for low-dissipation spintronic applications. In magnetoelectric materials, electric fields provide an additional control mechanism through spin–polarization coupling. Here we investigate, using classical Monte Carlo simulations, the combined effects of magnetic and electric fields on skyrmion phases in a ferromagnetic Heisenberg model on the square lattice with Dzyaloshinskii–Moriya interaction and magnetoelectric coupling via the
d-p hybridization mechanism. We analyze spin and dipolar textures, structure factors, magnetization, polarization, and scalar chirality for different field orientations and strengths, identifying ferromagnetic, ferroelectric, spiral, skyrmion crystal, skyrmion gas, and bimeron phases, as well as the field-driven transitions between them. We show that electric fields strongly reshape the stability region and internal structure of chiral phases, inducing skyrmion deformation, transmutation into bimerons, and shifts of the chiral window in magnetic field. Concomitant changes in magnetization and polarization across phase boundaries reflect the intrinsic magnetoelectric coupling characteristic of type-II multiferroics. Our results highlight the role of localized magnetoelectric entities, such as skyrmions carrying electric quadrupolar textures, in mediating electric-field control of topological magnetic states, providing a microscopic framework relevant to multiferroic skyrmion-host materials.

\end{abstract}

\maketitle

\section{\label{sec:level1}Introduction}

Magnetic skyrmions were theoretically predicted in the past century \cite{Bogdanov1989, Bogdanov1994} and first observed in the itinerant ferromagnet MnSi \cite{Binz2006, Muhlbauer2009}. Since then, they have been found in a wide variety of materials \cite{Neubauer2009, Munzer2010, Yu2010, Yu2011, Yu2012, Woo2016, Karube2018}. Due to their stability and small size, skyrmions are serious candidates for use in information storage and processing \cite{Nagaosa2013, Fert2013, Back2020, Gobel2021, Psaroudaki2021, Psaroudaki2023, Xia2023}, and they have therefore been the focus of numerous studies in recent years \cite{Pappas2009, Benerjee2014, Yi2009, Buhrandt2013, Huang2012, Romming2013, Romming2015, Iwasaki2013, Hayami2022, Heinze2011}.
These topologically nontrivial spin textures are typically stabilized by the Dzyaloshinskii-Moriya interaction (DMI) \cite{Dzyaloshinskii1958, Moriya1960} in the presence of an external magnetic field. However, alternative stabilization mechanisms have also been identified, including magnetic frustration \cite{Okubo2012,Leonov2015,Mohylna2022}, anisotropic couplings \cite{Gao2020,Rosales2022,Amoroso2020}, Ruderman-Kittel-Kasuya-Yosida (RKKY) \cite{Wang2020}, and dipolar \cite{Utesov2021} interactions. 

Skyrmions can be manipulated in metallic systems by ultralow current densities via spin-transfer torque \cite{Berger1996, Slonczewski1996, Jonietz2010, Schulz2012, Everschor2011, Zang2011}, although Ohmic heating remains a major limitation for spintronic applications. This has motivated increasing interest in magnetoelectric (ME) materials, particularly type-II multiferroics, where the electric polarization is induced by the magnetic order itself. In these systems, magnetization and polarization are intrinsically coupled, enabling the control of magnetic textures by electric fields with reduced energy dissipation \cite{Kimura2003, Fiebig2005, Tokura2006, Kimura2007, Cheong2007, Tokura2010, Khomskii2009} and providing a direct experimental signature of magnetoelectric phase transitions \cite{Ruff2015}.

The microscopic origin of this coupling is commonly described in terms of three main mechanisms \cite{Jia2006, Jia2007, Murakawa2010}. Two of them arise from correlations between neighboring spins. In noncollinear magnets, the spin-current mechanism generates a polarization of the form $\Vec{P}\propto \sum_{ij}\Vec{e}_{ij}\times(\Vec{S}_i\times\Vec{S}_j)$ \cite{Katsura2005, Mostovoy2006, Sergienko2006}. In systems with inequivalent magnetic sites, symmetric exchange striction can induce ferroelectricity through terms proportional to $\Vec{S}_i\cdot\Vec{S}_j$, allowing polarization even in collinear states \cite{Arima2006, Choi2008, Tokunaga2008, Ishiwata2010, Cabra2025}.

A third mechanism, particularly relevant in multiferroic insulators, is the spin-dependent $d$--$p$ hybridization between transition-metal and ligand ions \cite{Jia2006, Jia2007, Arima2007, Murakawa2012, Seki2012a, Liu2013a, Liu2013b}. Unlike the previous two, this mechanism is essentially local, with the induced dipole depending on the orientation of each spin relative to the bond direction, $\Vec{p}_{ij}\propto (\Vec{e}_{ij}\cdot\Vec{S}_i)^2\Vec{e}_{ij}$, and successfully accounts for magnetoelectricity in compounds such as CuFeO$_2$ \cite{Kimura2006}, CuCrO$_2$ \cite{Seki2008}, Ba$_2$CoGe$_2$O$_7$ \cite{Murakawa2010}, and Cu$_2$OSeO$_3$, where electric polarization emerges from the spin texture itself \cite{Belesi2012, Seki2012a, Seki2012b, White2012, White2014, Mochizuki2015a, Mochizuki2015b, Mochizuki2016, Okamura2016, Kruchkov2017}. 

In particular, Seki \textit{et al.} \cite{Seki2012a} showed that skyrmions in Cu$_2$OSeO$_3$ carry localized electric dipolar or quadrupolar moments, enabling their manipulation by external electric fields.  
These multipolar moments are not independent degrees of freedom, but are fully determined by the underlying spin texture through the magnetoelectric coupling. In this sense, dipolar and quadrupolar patterns provide a complementary representation of the same magnetic configurations, and their evolution reflects the deformation and reorganization of the skyrmionic structures.

In this work, we perform Monte Carlo simulations of a prototypical ferromagnetic skyrmion model incorporating a ME coupling \cite{Murakawa2012, Seki2012b}. In contrast to previous studies that focused on fixed magnetic or electric fields, here we provide a systematic mapping of the combined
(B,E) phase diagrams for two distinct electric-field orientations. This approach reveals how electric fields not only suppress or deform skyrmion lattices, but also induce controlled transformations between skyrmion, skyrmion-gas, and bimeron-rich phases across extended regions of parameter space.

\section{Model and methods}\label{sec:Model and Methods}

\begin{figure}[t!]
    \centering
    \includegraphics[width=0.4\textwidth]{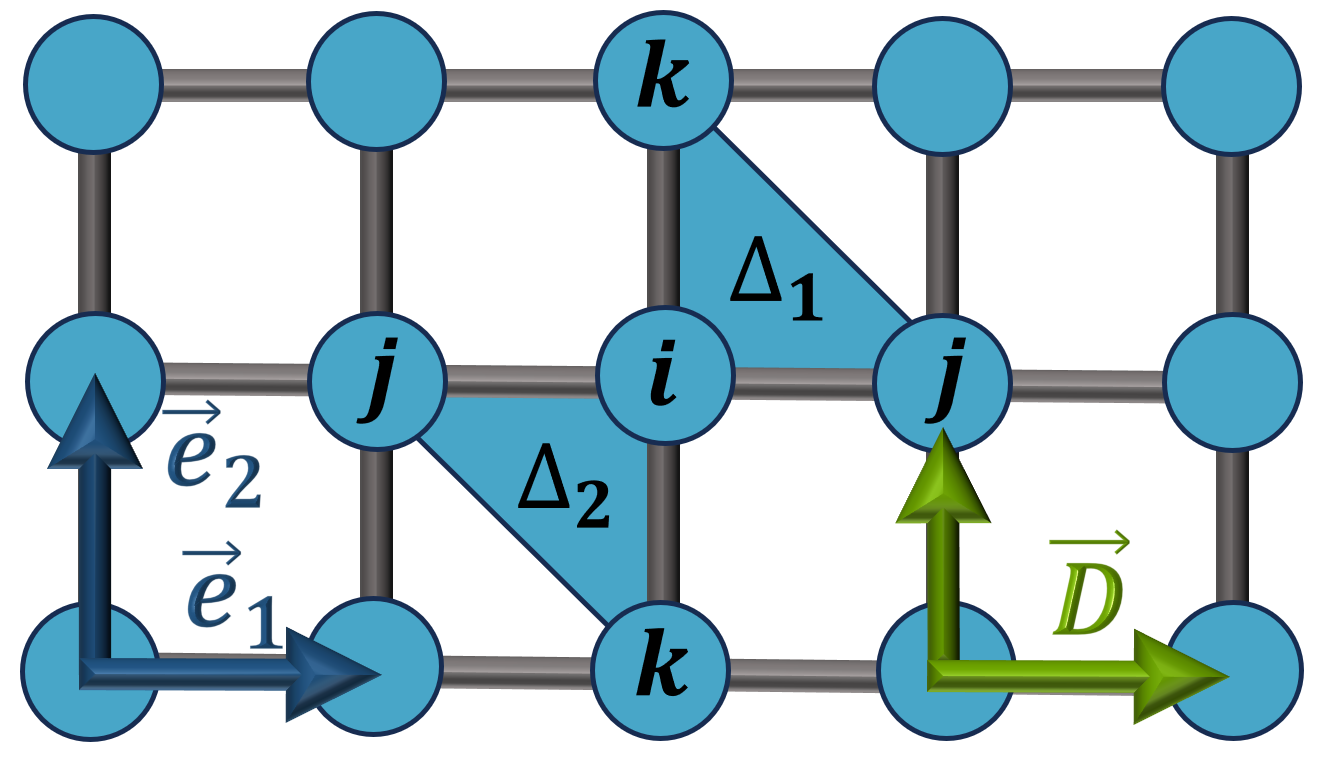}
    \caption{Square lattice scheme. The blue arrows represent the primitive translation vectors $\Vec{e}_1 = (1,0)$ and $\Vec{e}_2 = (0,1)$, the green arrows are the DM vectors $\Vec{D} = D\delta\hat{\mathrm{r}}$, and the labels $i, j, k$ indicate the sites involved in the calculation of the local chirality in the triangles $\Delta_1$ and $\Delta_2$.}
    \label{Fig0}
\end{figure}

We considered here a system composed of classical Heisenberg magnetic moments, in a square under uniform magnetic ($B_z$) and electric ($\Vec{E}$) fields, given by the Hamiltonian:
\begin{equation}
	\begin{split}
    \mathcal{H}  =  & - J \sum_{\langle i,j \rangle} \Vec{S}_i\cdot\Vec{S}_j + D \sum_{\langle i,j \rangle} 											\delta\hat{\mathrm{r}} \cdot (\Vec{S}_i \times \Vec{S}_j)\\
     				&    - \Vec{E} \cdot \sum_{i} \Vec{P}_i  -  B_z \sum_{i} S^z_i
    \label{1}
    \end{split}
\end{equation}
where the spin variables are unimodular classical vectors, $|\Vec{S}_i|=1$, $\Vec{P_i}$ is the electric dipole moment, $J > 0$ is the ferromagnetic exchange coupling, $D$ is the DM coupling, $\delta\hat{\mathrm{r}} = (\Vec{\mathrm{r}}_j-\Vec{\mathrm{r}}_i)/|\Vec{\mathrm{r}}_j-\Vec{\mathrm{r}}_i|$ is a unitary vector pointing along the axis, $\langle i,j \rangle$ indicates nearest neighbors (NN) coupling and the primitive translation vectors are $\Vec{e}_1 = (1,0)$ and $\Vec{e}_2 = (0,1)$ (Fig.\hspace{0.5mm}\ref{Fig0}).

\begin{figure}[t!]
    \centering
    \includegraphics[width=0.48\textwidth]{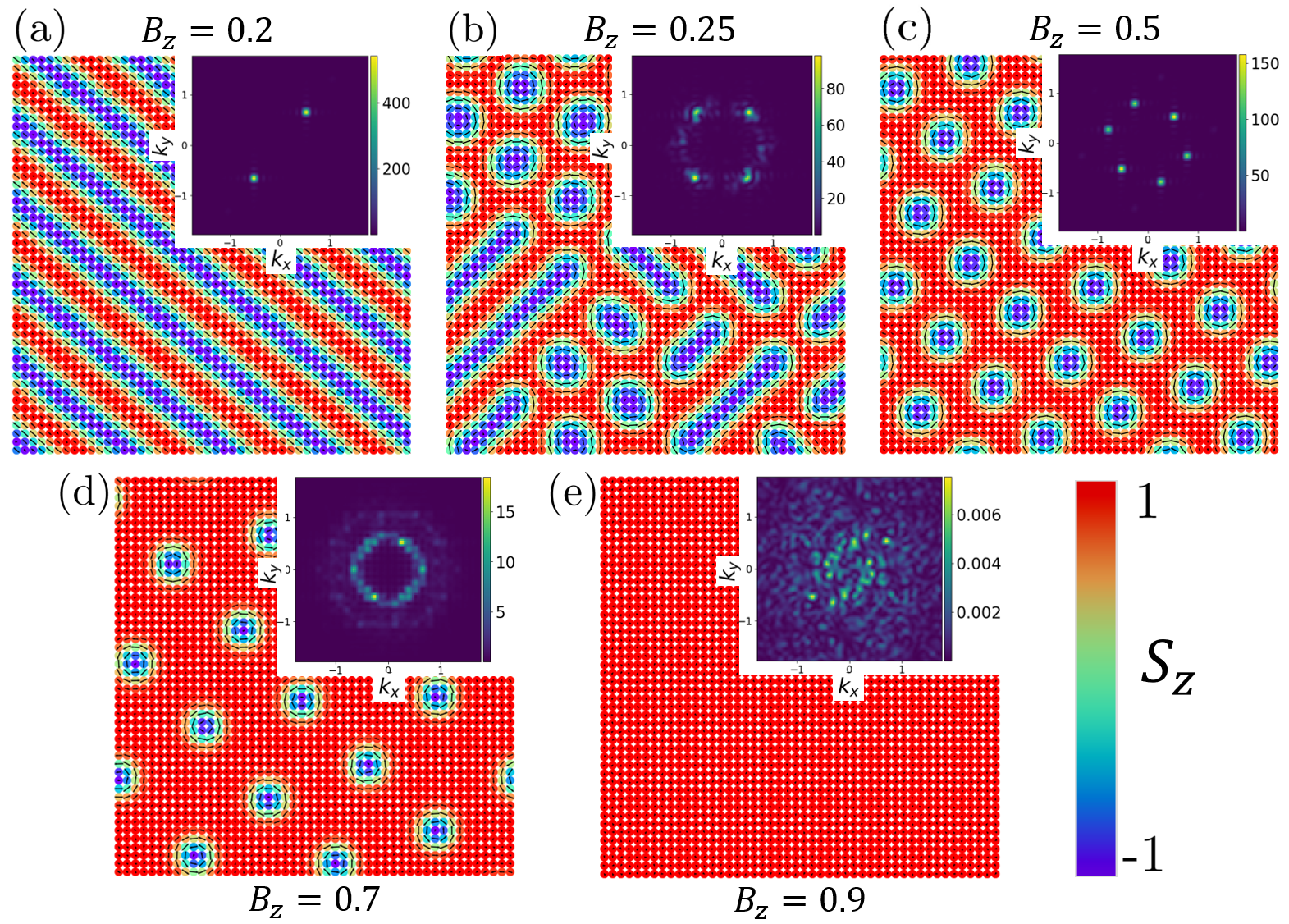}
    \caption{Spin textures and perpendicular (to z) structure factors ($\Vec{S}_{\perp}(\Vec{q})$) obtained at $E=0$ for representative values of $B_z$, showing the typical low-temperature phases: (a) Sp, (b) Bm, (c) SkX, (d) SkG, and (e) FM. The corresponding Bragg peaks in $\Vec{S}_{\perp}(\Vec{q})$ characterize the ordering in each phase.}
    \label{Fig1}
\end{figure}

In the previous section we have introduced some of the proposed mechanisms to explain the magneto-electric coupling. We considered here the spin-dependent metal–ligand hybridization mechanism, which has been shown to describe the magnetoelectric response in chiral magnetic insulators such as Cu$_2$OSeO$_3$ \cite{Seki2012a}. In this framework, the local polarization arises from a sum over bond contributions of the form $(\vec{e}_{ij}\cdot \vec{S}_i)^2\vec{e}_i$. Upon averaging over the bond directions $e_{ij}$ and considering the symmetry of the lattice, this expression reduces to an effective quadratic form in the spin components:

\begin{equation}
    \Vec{P_i} = \lambda (S^y_i S^z_i, S^z_i S^x_i, S^x_i S^y_i),
    \label{2}
\end{equation}
where $\lambda$ (set as $0.2$ in this work) is the magnetoelectric coupling constant. Setting $J=1$ fixes the energy scale, so that both $\vec{E}$ and $\lambda$ are dimensionless and the ME coupling is governed by $\lambda \vec{E}$. In this model the induced polarization in collinear states depends on the magnetization direction and may vanish along high-symmetry axes. As a result, the electric field favors configurations with mixed spin components, promoting noncollinear textures that can gain energy through the ME coupling.

\begin{figure*}[t!]
    \centering
    \includegraphics[width=1\textwidth]{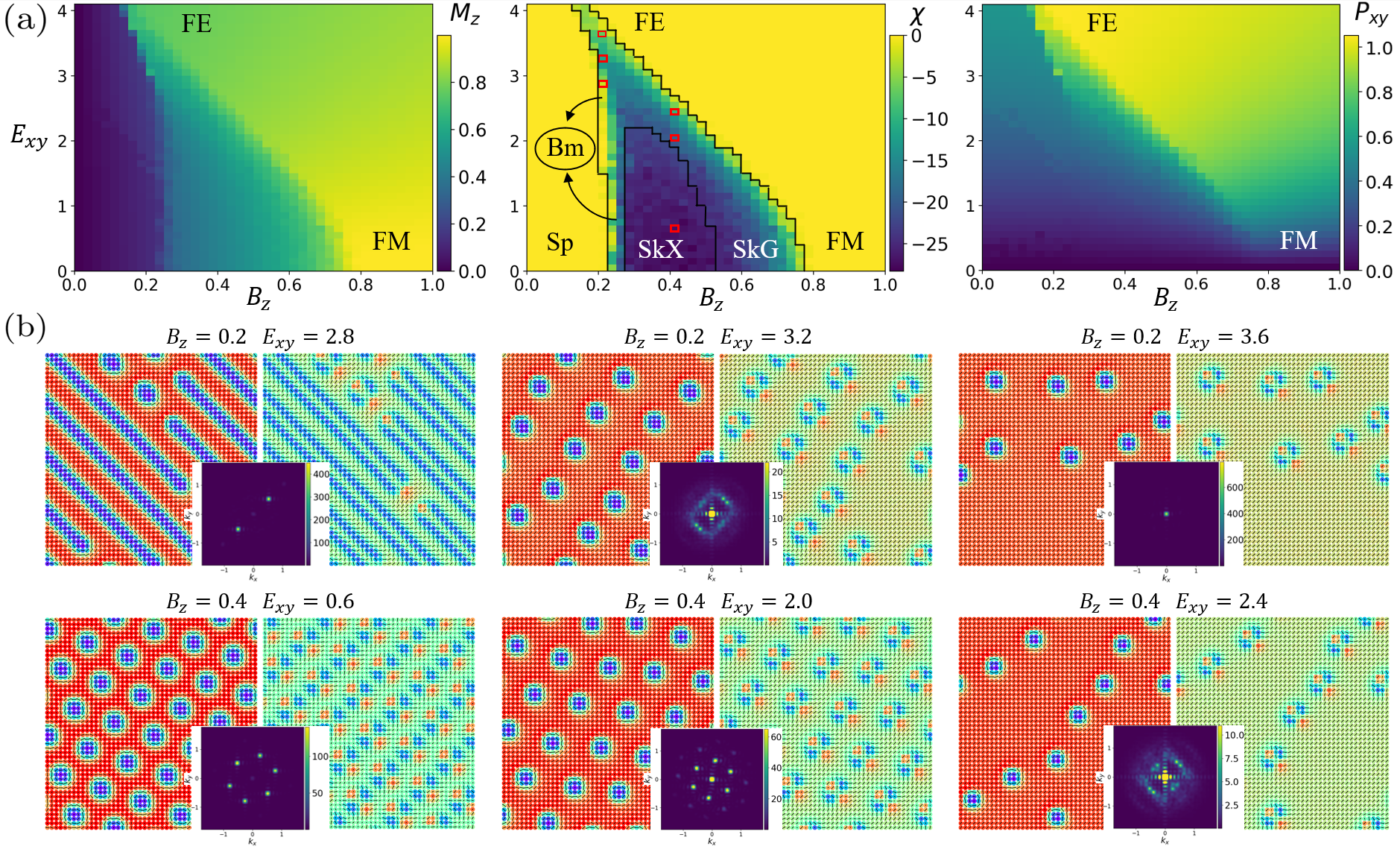}
    \caption{(a) Magnetization ($M_z$), chirality ($\chi$), and polarization parallel to the electric field ($P_{xy}$) are shown as functions of the fields $B_z$ and $E_{xy}$ in the square lattice, for $J=1$ and $D=1$ and $\mathrm{T} = 0.0009$. Black lines in the chirality are the boundaries between phases determined from $\chi$ and $\Vec{S_\perp}(\Vec{q})$. The red rectangles mark the position of the textures presented in the second part of this figure. (b) Representative spin and dipolar moment textures and structure factors for $B_z = 0.2$ and $B_{z} = 0.4$.}
    \label{Fig2}
\end{figure*}

We performed Monte Carlo (MC) simulations using the Metropolis algorithm on square lattices of size $N=L^2$ with $L=48$ and periodic boundary conditions. For each simulation, we lowered the temperature at a rate T$_{n+1}$ = 0.9T$_n$, from $T=2$ to $T \approx 10^{-3}$. At each temperature, up to $10^5$ Monte Carlo steps (MCS) were used for thermalization, followed by $2\times10^5$ MCS for measurements.
To improve statistical reliability and assess metastability effects, we performed 10 independent runs for each set of parameters, increasing this number to 20 in regions close to phase transitions. In addition, complementary simulations were carried out at fixed low temperature ($T \approx 10^{-3}$), sweeping the electric field from $E=0$ to a maximum value and back, in order to assess the stability of the observed phases.
To evaluate finite-size effects, we also performed 5 independent simulations for larger system sizes ($L=60$ and $L=72$) at selected points of the phase diagram, confirming that the main features remain qualitatively unchanged.

The different phases are identified through a combined analysis of real-space spin textures, structure factors, and thermodynamic observables. We have calculated the perpendicular (to z) component of the static spin structure factor in the reciprocal lattice ($\Vec{S}_{\perp}(\Vec{q})$) to identify the Bragg peaks that characterize the different spin-textures. It is defined as:

\begin{equation}
    \Vec{S}_{\perp}(\Vec{q}) = \frac{1}{N} \langle |\sum_{j}  S_j^x  e^{i \Vec{q} \cdot \hat{r}_{j}}|^{2} +|\sum_{j}  S_j^y  e^{i \Vec{q} \cdot \hat{r}_{j}}|^{2} \rangle,
    \label{4}
\end{equation}

where $\langle \rangle$ means the thermal average.

We computed the average values of the magnetization, specific heat, and susceptibility, and, since the model (\ref{1}) may present skyrmionic phases, the total scalar spin chirality (discrete topological charge), defined as:
\begin{equation}
    \chi = \left\langle \frac{1}{4\pi} \sum_{\Delta_1,\Delta_2} \Vec{S}_i \cdot (\Vec{S}_j \times \Vec{S}_k) \right\rangle, 
    \label{3} 
\end{equation}
where $\Delta_1$ and $\Delta_2$ run over all the elementary triangles formed by NN spins in sites $i$, $j$ and $k$ respectively. On the square lattice, each plaquette is decomposed into two triangles, and a consistent counterclockwise ordering of the spins is adopted (see Fig.\hspace{0.5mm}\ref{Fig0}). This convention fixes the sign of the scalar chirality and therefore determines the sign of the associated topological charge, which reflects the rotational sense (chirality) of the skyrmionic textures. In the parameter regime explored here (fixed signs of $J$, $D$, and $B_z > 0$ ), the skyrmion chirality is positive, and the electric field is not expected to alter it, as the magnetoelectric term is quadratic in the spin components.

Chirality provides a direct measure of the number of skyrmions, as it counts the total winding of the spin texture; each skyrmion carries an approximately quantized charge $Q\approx1$, so that the total charge is proportional to their number. When convenient, we normalize this quantity by plotting ${\chi}/{N_S}$, where $N_S$ is the maximum number of skyrmions attained upon cooling, or ${\chi}/{L^2}$, where $L^2$ is the number of spins in the lattice.

\begin{figure}[t!]
    \centering
    \includegraphics[width=0.45\textwidth]{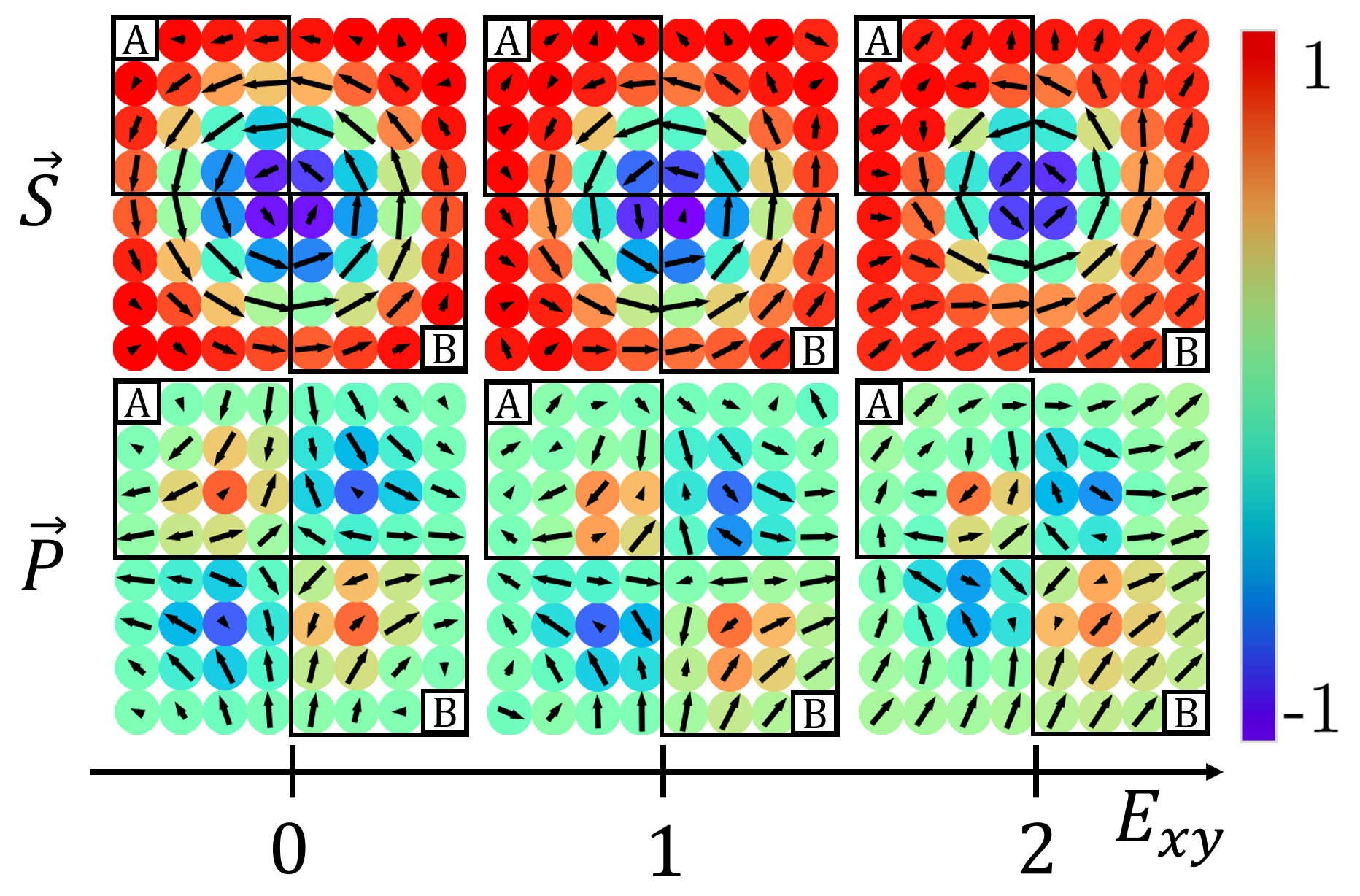}
    \caption{Spin and dipole moment textures of isolated skyrmions for fields $B_z = 0.5$ and $E_{xy} = 0, 1, 2$. A and B squares are used to highlight the behavior of applying an electric field.}
    \label{Fig3}
\end{figure}

\begin{figure}[t!]
    \centering
    \includegraphics[width=0.45\textwidth]{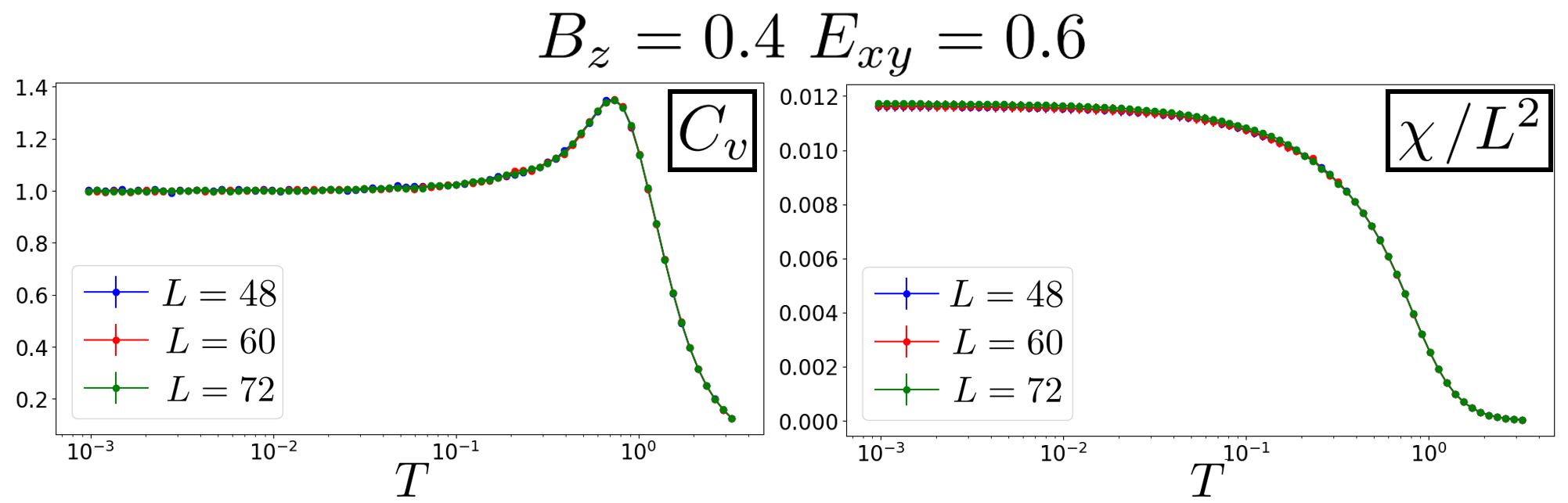}
    \caption{Specific heat and chirality per site for $B_z=0.4$ and $E_{xy}=0.6$ for lattice sizes L = 48, 60, 72.}
    \label{Fig3b}
\end{figure}

Throughout this work we set $J=1$, which fixes the energy scale, and $D=1$, known to stabilize skyrmions. While the precise phase boundaries depend on the ratio $D/J$, the qualitative features reported here are expected to remain valid in the regime where exchange and DM interactions are comparable. We first consider the well-known case $E=0$ on the square lattice  \cite{Albarracin2022}. Fig.\hspace{0.5mm}\ref{Fig1} shows the spin textures at the lowest temperature reached (T=0.0009). The upper right corner displays the first Brillouin zone of the perpendicular (to z) structure factor. In (a), for magnetic field $B_z = 0.2$, the helical or spiral phase (Sp) is observed, characterized by its two peaks structure factor, known as single-q \cite{Gobel2021}. In (c), for $B_z = 0.5$, the skyrmion crystal phase (SkX) emerges. Its structure factor, known as triple-q, displays six peaks as consequence of the superposition of three spiral phases.  In (e), for $B_z = 0.9$, the ferromagnetic (FM) phase is reached, where spins align along the magnetic field direction and $\Vec{S}_{\perp}(\Vec{q}) \approx \vec{0}$ \hspace{0.1mm} $\forall \vec{q}$.

Ezawa et al.\cite{Ezawa2011} analyzed the free energy of the Sp and SkX phases and showed that, at T=0, a critical magnetic field favors the complete breakup of spiral state into a skyrmion lattice. However, at any finite temperature, a gap opens at the Sp–SkX phase transition, in which excitations (merons) are stabilized, partially disrupting the spiral order and giving rise to the bimeron (Bm) phase, or “elongated skyrmions" (depicted in panel (b)). As the spiral phase breaks down, the single-q structure factor is also lost, since bimerons can orient along both diagonal directions of the lattice, leading to a double-q ordering pattern.
A similar argument applies to the SkX–FM transition, where thermal fluctuations  stabilizes a skyrmion gas (SkG) phase, whose structure factor exhibits a ring-like pattern due to the absence of a preferred orientation, as shown in panel (d).

\begin{figure*}[t!]
    \centering
    \includegraphics[width=0.98\textwidth]{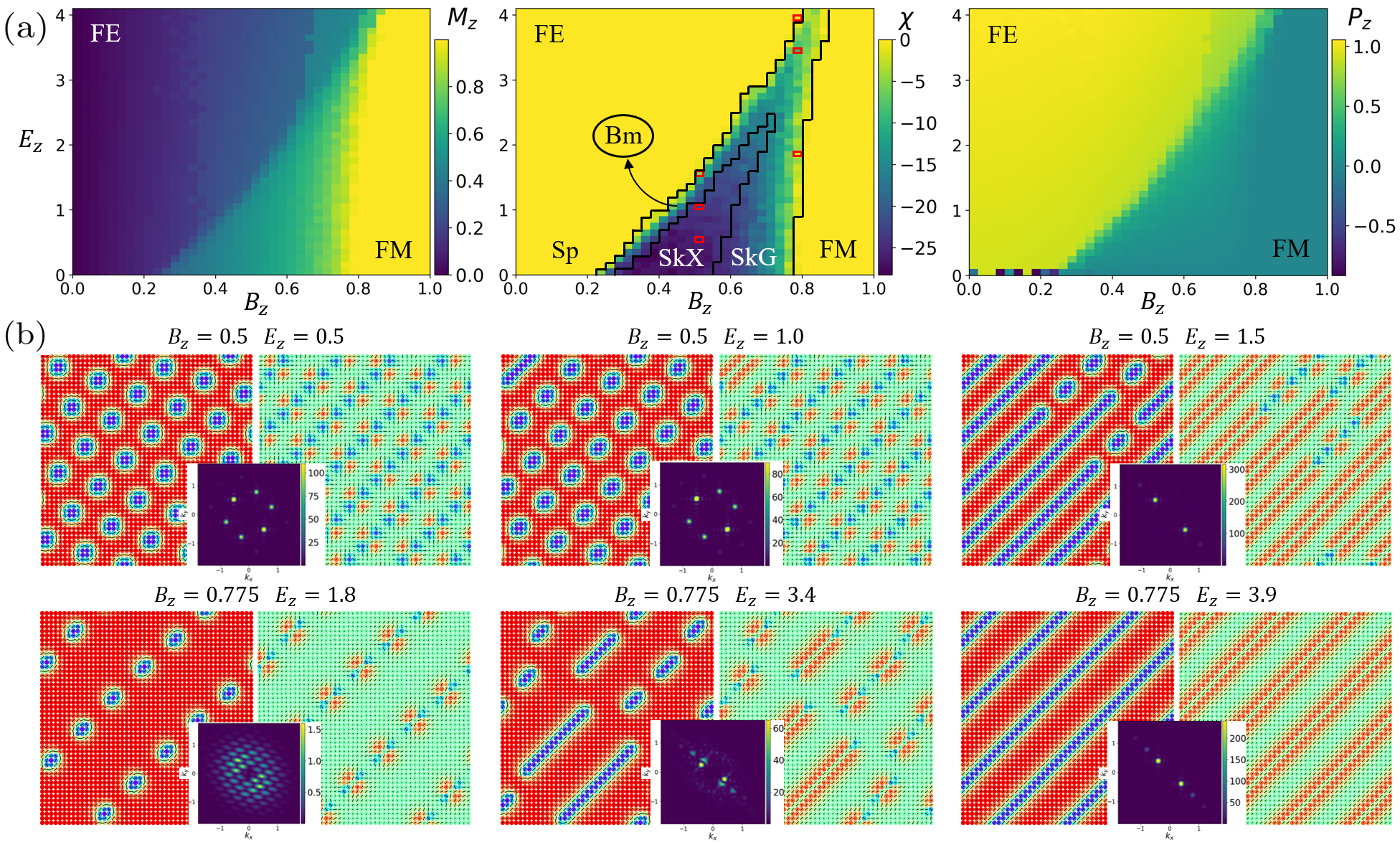}
    \caption{(a) Magnetization ($M_z$), chirality ($\chi$), and polarization parallel to the electric field ($P_{z}$) are shown as functions of the fields $B_z$ and $E_{z}$ in the square lattice, for $J=1$ and $D=1$ and $\mathrm{T} = 0.0009$. Black lines in the chirality are the boundaries between phases determined from $\chi$ and $\Vec{S_\perp}(\Vec{q})$. The red rectangles mark the position of the textures presented in the second part of this figure. (b) Representative spin and dipolar moment textures and structure factors for $B_z = 0.5$ and $B_z = 0.775$.}
    \label{Fig4}
\end{figure*}

\section{Low-temperature behaviour}

In this section, we analyze the results obtained from MC simulations at low temperature. Subsection A presents and discusses the most relevant aspects of the phase diagrams for the two electric field directions, together with representative spin and dipole moment textures and their corresponding structure factors. We also examine the deformation of skyrmions under the action of the electric field. Subsection B focuses on the thermodynamic variables as functions of one of the fields, while fixing the other at characteristic values.

\subsection{Phase diagrams in $(B,E)$}

\subsubsection{$\Vec{E} \parallel (1,1,0)$}

In Fig.\hspace{0.5mm}\ref{Fig2}(a) we show the magnetization, chirality, and polarization parallel to the electric field as functions of $B_z$ and $E_x = E_y = E_{xy}$, with $E_z=0$, at $T \approx 10^{-3}$. From the magnetization and xy-polarization, we identify the competition between the magnetic and electric fields in aligning the spins. At low $E_{xy}$ and high $B_z$, the system is ferromagnetic (FM). In contrast, at high $E_{xy}$ and low $B_z$, the spins align in the $(1/2, 1/2, 1/\sqrt{2})$ direction, as can be derived from Eqs.~(\ref{1}) and (\ref{2}), minimizing the Hamiltonian per spin ($h \approx -(S^y S^z+S^x S^z)E_{xy}$). This direction maximizes the polarization component parallel to the electric field, giving rise to a ferroelectric ($\mathrm{FE}$) phase.

The chirality map depicts the complete phase diagram, with phase boundaries determined from chirality and structure factors. In the intermediate region, where skyrmions are present at zero electric field, applying $E_{xy} \approx 3$ destroys the skyrmions, although intermediate deformations appear, analyzed below in this subsection. To illustrate this behavior, spin and dipole moment textures, together with their structure factors for $B_z = 0.4$, are shown in Fig.\hspace{0.5mm}\ref{Fig2}(b). The triple-$q$ structure of the SkX phase, well defined at $E_{xy}=0.6$, exhibits a central peak (parallel spins with nonlinear xy components) accompanied by secondary maxima. At $E_{xy}=2.0$, the lattice begins to separate. At $E_{xy}=2.4$, the dispersion is complete and the central peak sharpens, corresponding to a SkG phase with a parallel-spin background. 

Notably, the competition between the fields also shifts the chiral region to lower magnetic fields for $E_{xy} \approx 3$. This is evidenced by the textures and structure factors in Fig.\hspace{0.5mm}\ref{Fig2}(b). At $B_z = 0.2$, increasing $E_{xy}$ up to $2.8$ destabilizes the spiral phase, leading to a mixed state of skyrmions and bimerons. The corresponding single-$q$ structure factor develops secondary intensity maxima. The number of skyrmions increases up to a maximum at $E_{xy} = 3.2$, where a skyrmion gas forms, characterized by a ring-like distribution of $\Vec{S}_{\perp}$. For $E_{xy} \geq 3.8$, the number of skyrmions decreases, giving way to the parallel-spin phase, whose $\Vec{S}_{\perp}$ shows a sharp peak at the origin due to spin alignment with a finite xy component, as also observed at $E_{xy}=3.6$.
This result is particularly significant, as it demonstrates how the application of an electric field can effectively tune the chiral region, enabling the stabilization of skyrmions at lower magnetic fields. Such electric-field control over the stability and extent of chiral phases highlights a key mechanism for experimentally manipulating skyrmion-hosting states in magnetoelectric systems.

Beyond the overall composition of the phase diagram, the electric field strongly affects the skyrmion shape. In Fig.\hspace{0.5mm}\ref{Fig3}, we show spin and dipole moment textures for a single skyrmion at $B_z=0.5$ and $E_{xy}=0,1,2$. The electric field competes with the magnetic field and DMI, progressively disassembling the skyrmion. Spins in region B tilt toward the xy-plane as $E_{xy}$ increases, while those in region A tilt toward the z direction. At $E_{xy} = 2$, the region B starts merging with the background. 
The skyrmion–quadrupole relationship allows us to interpret this deformation more directly in dipolar-moment space: the dipoles align with the electric field such that in region B the quadrupole increases in size but decreases in intensity as the $P_z$ component diminishes, while in region A the area with $P_z > 0$ is reduced. 

To assess possible finite-size effects, we performed additional simulations for different lattice sizes ($L=48, 60, 72$) under the same sets of external fields. These comparisons were carried out at selected points of the phase diagram, chosen to be representative of the different phases. The results indicate that the main features of the phase diagram remain largely unchanged as the system size increases. 
This behavior is illustrated in Fig.\hspace{0.5mm}\ref{Fig3b}, where we show the case $E_{xy}=0.6$ and $B_z=0.4$, for which the low-temperature phase corresponds to a skyrmion lattice. The thermodynamic observables displayed (specific heat and chirality) exhibit a consistent behavior for $L=48, 60, 72$. These findings suggest that the observed phases and transitions are not dominated by finite-size effects.

\begin{figure}[t!]
    \centering
    \includegraphics[width=0.45\textwidth]{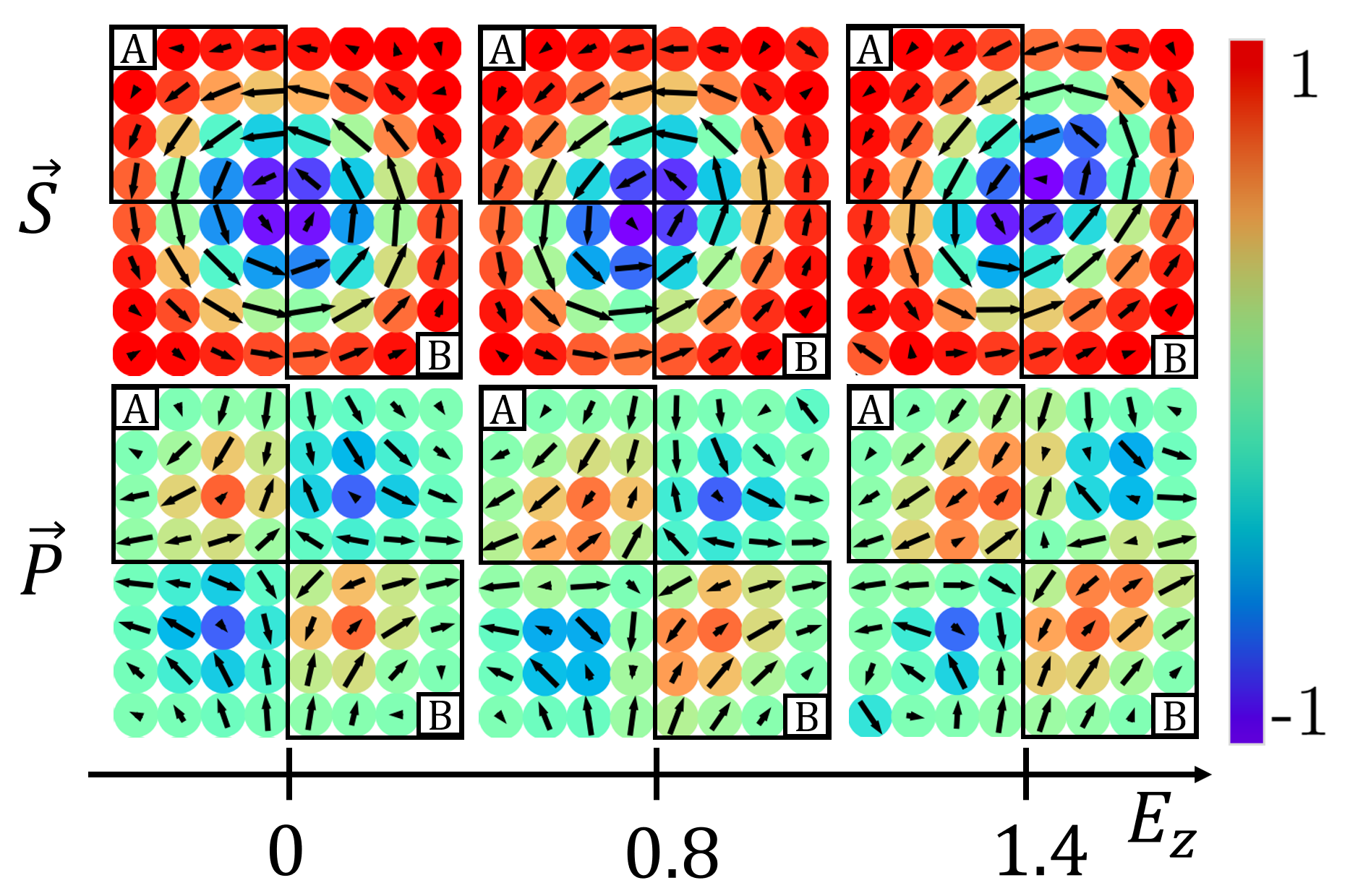}
    \caption{Spin and dipole moment textures of isolated skyrmions for fields $B_z = 0.5$ and $E_{z} = 0, 0.8, 1.4$. A and B squares are used to highlight the behavior of applying an electric field.}
    \label{Fig5}
\end{figure}

\begin{figure}[t!]
    \centering
    \includegraphics[width=0.45\textwidth]{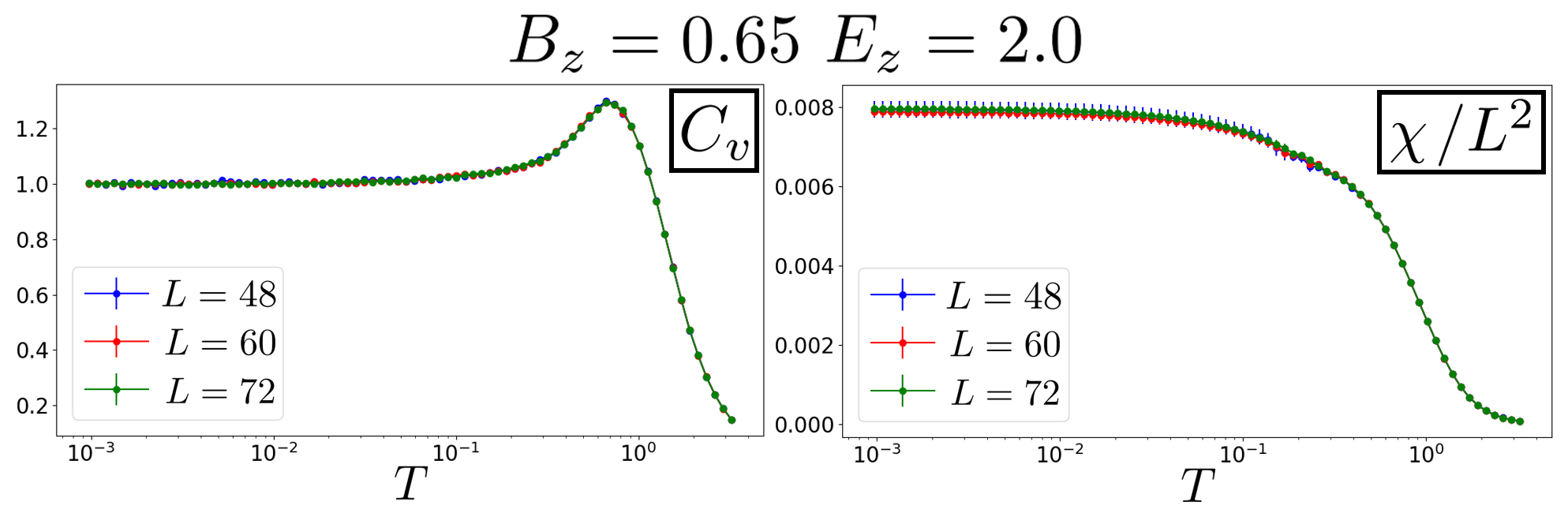}
    \caption{Specific heat and chirality per site for $B_z=0.65$ and $E_z=2.0$ for lattice sizes L = 48, 60, 72.}
    \label{Fig6b}
\end{figure}

\subsubsection{$\Vec{E} \parallel (0,0,1)$}

In Fig.\hspace{0.5mm}\ref{Fig4}(a), we present the magnetization, chirality, and z-polarization as functions of $B_z$ and $E_z$, with $E_x=E_y=0$, at $T \approx 10^{-3}$. As in the previous case, the magnetization and polarization reflect the competition between magnetic and electric fields. At high $B_z$ and low $E_z$, the system is ferromagnetic (FM). Conversely, at low $B_z$ and high $E_z$, the spins align along $(1/\sqrt{2},1/\sqrt{2},0)$ with $\vec{P}=(0,0,1)$, consistent with Eqs.~(\ref{1}) and (\ref{2}), minimizing the Hamiltonian when $E_z \gg B_z, |J|, D$.

The phase diagram, obtained from chirality and structure factors, is shown in the $\chi$ plot. The chiral region is destroyed by sufficiently strong electric fields, with the critical $E_z$ decreasing at lower $B_z$. For example, Fig.\hspace{0.5mm}\ref{Fig4}(b) shows that the SkX phase observed at $B_z=0.5$, $E_z=0.5$ (with the characteristic triple-$q$ pattern in $\Vec{S}_\perp$) evolves into a spiral phase with a single-$q$ structure at $E_z=1.5$. 

\begin{figure*}[ht!]
    \centering
    \includegraphics[width=0.98\textwidth]{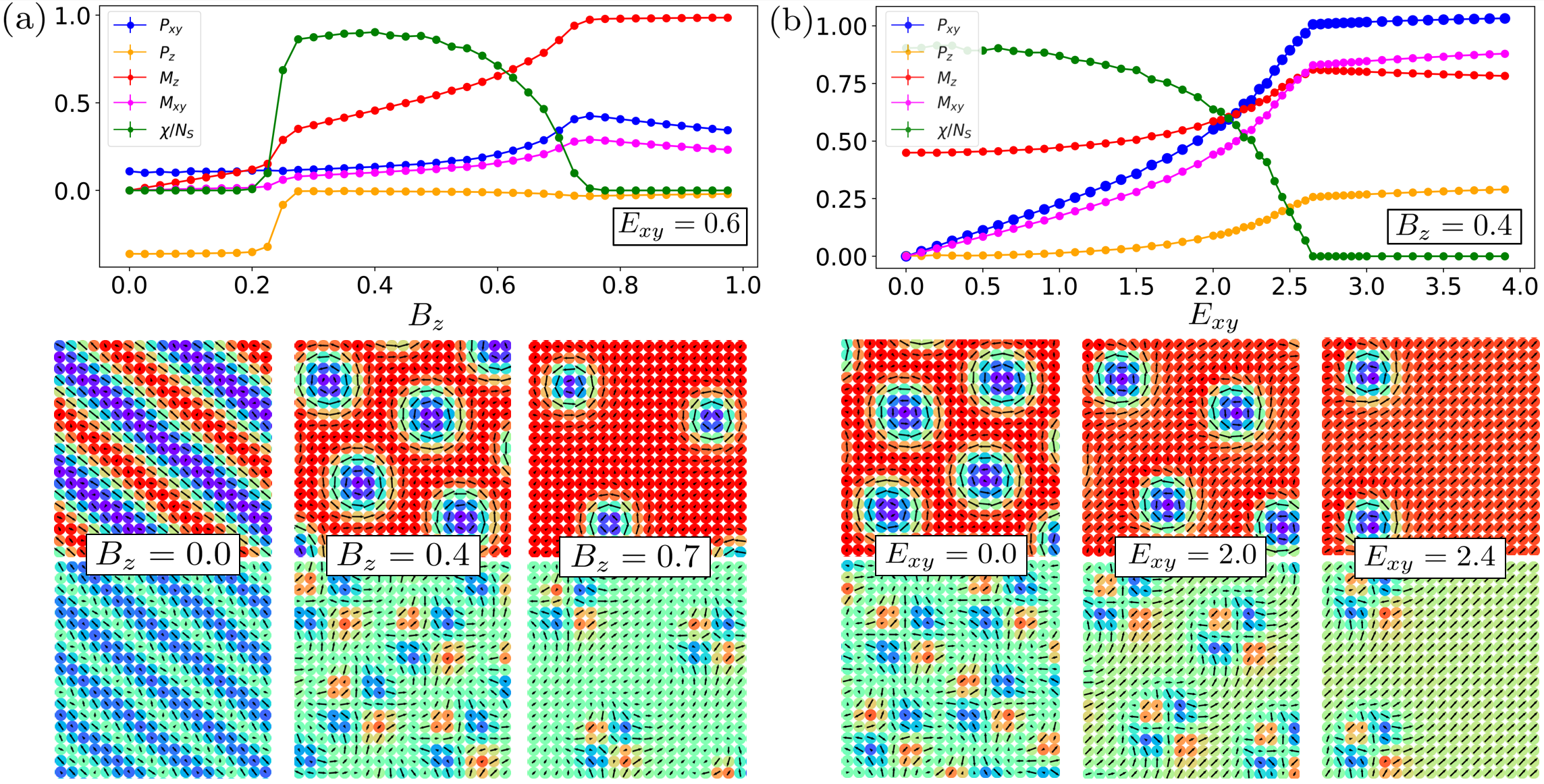}
    \caption{From top to bottom: xy-polarization, z-polarization, z-magnetization, xy-magnetization, and normalized chirality (with $N_S = 30$) at $T=0.0009$, as functions of $B_z$ for $E_{xy} = 0.6$ in (a), and as functions of $E_{xy}$ for $B_z = 0.4$ in (b). Error bars are smaller than the marker size when not visible. Below: representative spin and dipole moment textures for each case.}
    \label{Fig7}
\end{figure*}

\begin{figure*}[ht!]
    \centering
    \includegraphics[width=0.98\textwidth]{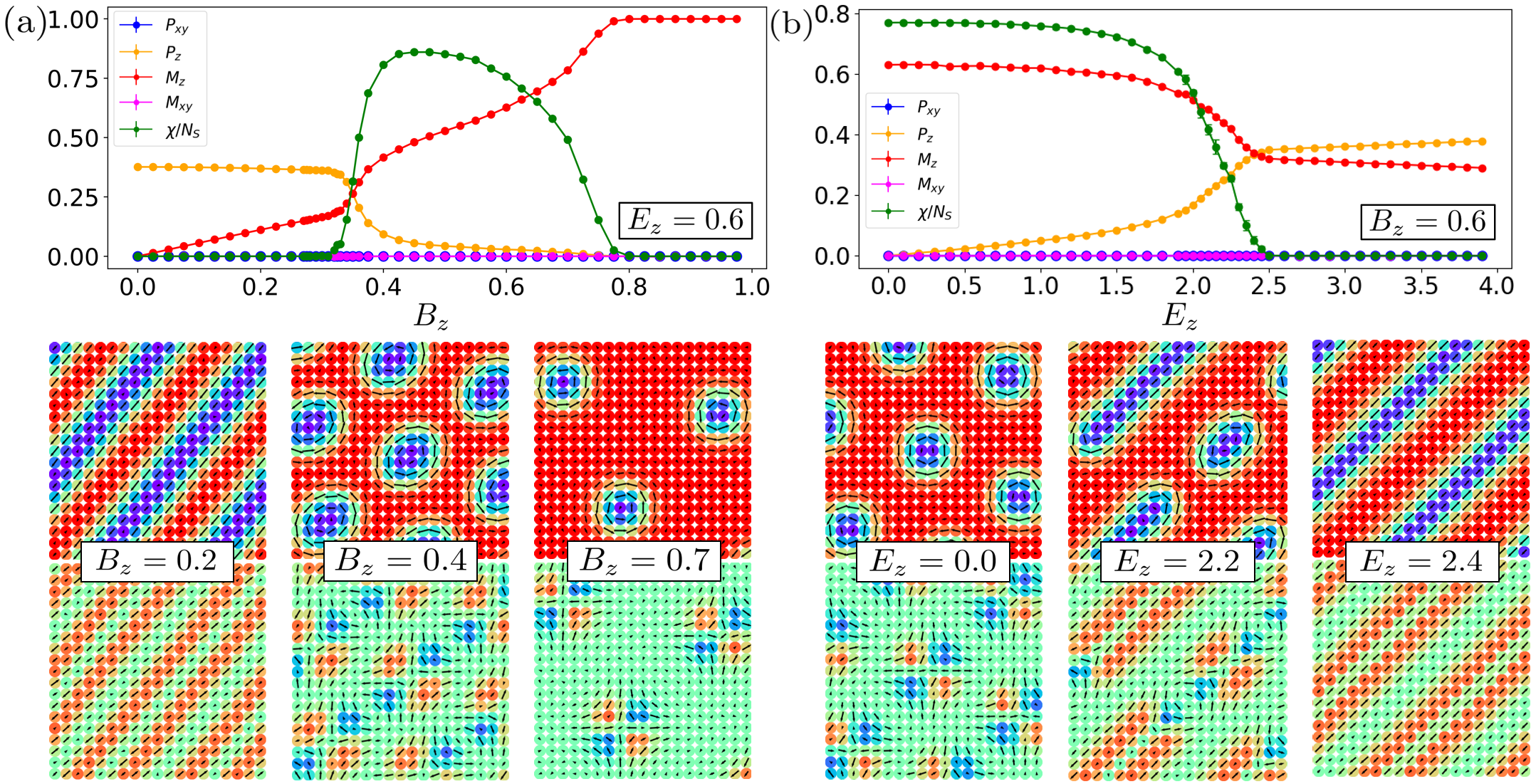}
    \caption{From top to bottom: xy-polarization, z-polarization, z-magnetization, xy-magnetization, and normalized chirality (with $N_S = 30$) at $T=0.0009$, as functions of $B_z$ for $E_{z} = 0.6$ in (a), and as functions of $E_{z}$ for $B_z = 0.6$ in (b). Error bars are smaller than the marker size when not visible. Below: representative spin and dipole moment textures for each case.}
    \label{Fig8}
\end{figure*}

A related effect is observed at larger magnetic fields, around $B_z \approx 0.8$. In this region, the ground state at $E=0$ is ferromagnetic; however, upon increasing $E_z$, chiral phases extend into this field range. In other words, the application of an electric field stabilizes noncollinear textures in a parameter region that would otherwise be collinear. As shown in Fig.\hspace{0.5mm}\ref{Fig4}(b), for $B_z=0.775$ and $E_z \approx 2$ we obtain a SkG phase in which skyrmions elongate and partially align. The corresponding $\Vec{S}_\perp(\vec{q})$ shows a broadened ring with pronounced peaks, typical of elongated skyrmions. For $E_z \approx 3$, further elongation and asymmetry lead to a bimeron-rich (Bm) phase. At even larger $E_z$, spiral states dominate, with wide ferromagnetic regions favored by the high $B_z$. 
As in the previous case, this result provides evidence of electric manipulation of magnetic structures, enabling the creation, annihilation, and deformation of skyrmions in new regions of magnetic field.

Skyrmion deformation is analyzed in Fig.\hspace{0.5mm}\ref{Fig5}. Here, skyrmions elongate preferentially along the xy direction. This effect is more evident in the dipole moments: the z-component of the dipoles in regions A and B increases with $E_z$ due to magnetoelectric coupling, while dipoles with negative $P_z$, located along the xy-diagonal (outside A and B), rotate toward the xy-plane. 
This redistribution of dipolar moments, together with the skyrmion–quadrupole relationship, provides a clear interpretation of the skyrmion elongation and the previously described SkX–Bm–Sp and SkG–Bm–Sp phase transitions.

To further evaluate finite-size effects, we carried out additional simulations for lattice sizes $L=48, 60, 72$ under identical external conditions. These checks were performed at representative points of the phase diagram, covering different types of magnetic textures. The comparison shows that the overall behavior remains stable as the system size is increased.
An illustrative example is presented in Fig.\hspace{0.5mm}\ref{Fig6b}, corresponding to $E_z=2.0$ and $B_z=0.65$, where the low-temperature regime is characterized by a skyrmion lattice. Specific heat and chirality display a consistent evolution for the three system sizes. This indicates that the observed phase behavior and transition features are not significantly affected by finite-size effects.

\begin{figure*}[ht!]
    \centering
    \includegraphics[width=0.98\textwidth]{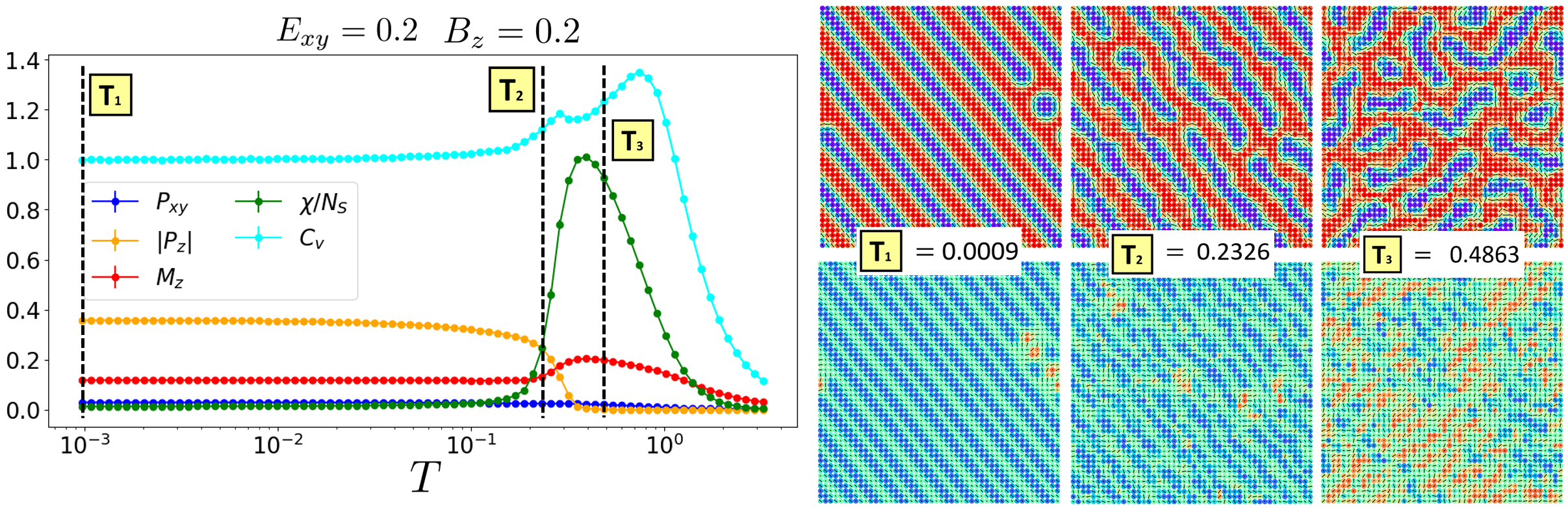}
    \caption{Left: temperature dependence of $z$ and $xy$ polarization, magnetization, normalized chirality (with $N_S = 12$), and specific heat for $E_{xy} = 0.2$ and $B_z = 0.2$. Error bars are smaller than the marker size when not visible. Black dashed lines mark the temperatures at which spin and dipole moment textures are displayed on the right.}
    \label{Fig11}
\end{figure*}

\subsection{ME effects in the order parameters}

To gain further insight into the magnetoelectric effects, we studied physical observables at low $T$ as functions of one field, while keeping the other fixed. In particular, we analyzed the scalar chirality defined in Eq.~(\ref{3}), the polarization components $P_z$ and $P_{xy} \equiv (P_x + P_y)/\sqrt{2}$, and the magnetization components $M_z$ and $M_{xy} \equiv (M_x + M_y)/\sqrt{2}$

We show the results as functions of $B_z$ for $E_{xy} = 0.6$ in Fig.\hspace{0.5mm}\ref{Fig7}(a). The chirality curve delineates the region where SkX and SkG phases are stabilized. Two clear phase boundaries are observed: $B_{c1} \approx 0.22$, marking the onset of the SkX phase, and $B_{c2} \approx 0.75$, where skyrmions vanish completely. These skyrmions are deformed, as previously discussed, a fact also reflected in $M_{xy}$, which acquires a small but finite value in the chiral region due to symmetry breaking. Remarkably, while the number of skyrmions decreases as $B_z$ increases, the background spins tend to align with the z axis, and yet $M_{xy}$ grows. This suggests that the remaining skyrmions continue to deform under stronger $B_z$. As expected, $M_z$ increases with $B_z$, with steeper slopes at $B_{c1}$ and $B_{c2}$. $P_z$ is negative in the spiral phase (blue textures at $B_z = 0$) and becomes very small at $B_{c1}$, where the SkX phase, consisting of subtly deformed skyrmions (and quadrupoles) due to the low $E_{xy}$ field, is stabilized. $P_{xy}$ follows a similar trend to $M_{xy}$, providing an additional indicator of ME coupling. 
In connection to potential experiments, it should be stressed here that there are concomitant changes in P and M at both phase boundaries, as is typically observed in type II multiferroics \cite{Cheong2007,Cabra2025} and references therein.
Figure~\ref{Fig7}(b) shows the evolution with $E_{xy}$ for fixed $B_z=0.4$. Here, $M_z$ increases across the chiral region and approaches asymptotically $1/\sqrt{2}$, as discussed earlier. Both $P_{xy}$ and $M_{xy}$ grow with $E_{xy}$, manifesting in spin textures as a rotation of the ferromagnetic background, concomitant with a reduction in skyrmion density.

The physical parameters for $\Vec{E} \parallel \hat{z}$ are presented in Fig.\hspace{0.5mm}\ref{Fig8}. For fixed $E_z=0.6$, $M_z$ grows nearly linearly with $B_z$, except for an abrupt increase at the helix–SkL transition, consistent with an enhancement of the ferromagnetic background. Interestingly, at $B_z \approx 3.4$, a crossover between $M_z$ and $|P_z|$ occurs, reminiscent of behaviors reported in multiferroic materials with strong ME coupling. This crossover reflects the replacement of spiral states, which favor $|P_z|>0$, by skyrmions and quadrupoles, whose dipolar distribution balances $P_z$. Similar behavior is observed at fixed $B_z=0.6$ as a function of $E_z$: $M_z$ decreases at the SkG–spiral transition (marked by a drop in chirality), while $|P_z|$ increases. As discussed in the previous subsection, these phase transitions —and the associated crossover between magnetization and polarization— are best understood through the combined analysis of the skyrmion and quadrupole spaces. Increasing the magnetic field $B_z$ promotes a more ferromagnetic background and balances the electric components, whereas increasing the electric field $E_z$ enhances $P_z$ and compensates the magnetic components of the system.

\begin{figure*}[t!]
    \centering
    \includegraphics[width=0.98\textwidth]{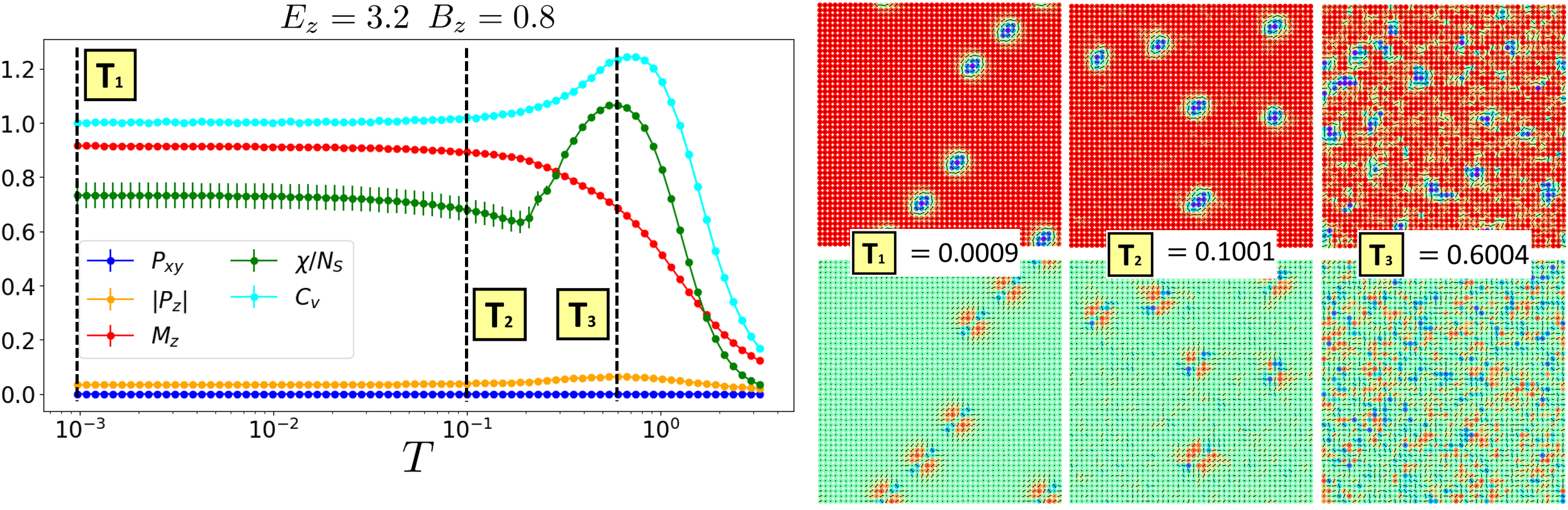}
    \caption{Left: temperature dependence of $z$ and $xy$ polarization, magnetization, normalized chirality (with $N_S = 6$), and specific heat for $E_{z} = 3.2$ and $B_z = 0.8$. Error bars are smaller than the marker size when not visible. Black dashed lines mark the temperatures at which spin and dipole moment textures are displayed on the right.}
    \label{Fig12}
\end{figure*}

\section{Thermal fluctuation effects}

Skyrmions in magnetic materials can persist at temperatures as high as $J$, which is one of the key features making them promising for spintronic applications. In this section, we present the most relevant results for the thermodynamic variables (magnetization ($M_{z}$ and $M_{xy}$), polarization ($P_z$ and $P_{xy}$), specific heat ($C_v$), and chirality ($X_M$)) as functions of temperature, while also discussing the associated dipolar textures.

Fig.\hspace{0.5mm}\ref{Fig11} shows the curves for $E_{xy} = 0.2$ and $B_z = 0.2$, together with the spin and dipole moment textures at $T \approx 10^{-3}$ compared with higher temperatures. At $T \approx 1$, where all variables vanish, the system reaches the paramagnetic phase. At $T \ll 1$, a low-chirality phase emerges, corresponding to a spiral state with an isolated skyrmion, as evidenced in the spin textures. 

During the cooling process, two intermediate chiral regimes appear due to thermal fluctuations. At $T \approx 0.5$, distorted and curved bimeron-like structures emerge, while at $T \approx 0.25$, these textures tend to align along one of the diagonal directions of the lattice, still affected by residual thermal fluctuations. These structural changes are reflected in the polarization $P_z$, which decreases abruptly, and in the specific heat $C_v$, which exhibits two peaks associated with these crossovers.
Thermal fluctuations induce the formation of elongated chiral structures which, on average, extend equally along both diagonal directions of the lattice above a characteristic temperature $T_c \approx 0.32$. Since $P_z$ is positive for structures oriented along $(1,1)$ and negative for those along $(-1,1)$, its net value cancels out, behaving as an effective indicator of this transition. This situation further highlights the relevance of analyzing the dipolar-moment space and the skyrmion–quadrupole relationship

For $E_z = 3.2$ and $B_z = 0.8$ (Fig.\hspace{0.5mm}\ref{Fig12}), the cooling process proceeds from the paramagnetic phase to a skyrmion gas (SkG). At intermediate temperatures, the number of skyrmions fluctuates slightly, and their size and shape vary (e.g., at $T=0.1001$). Another intermediate state emerges, characterized by small deformed skyrmions in larger numbers than at low temperatures, corresponding to a maximum in chirality. At higher temperatures, the skyrmion gas melts into a chiral disordered phase, reaching a maximum in the chirality at $T \approx 0.6$, where the spin textures exhibit smaller chiral structures. At low temperatures, the chirality exhibits noticeable error bars, which can be attributed to the presence of metastable configurations in some independent runs.

\section{CONCLUSIONS AND FUTURE PERSPECTIVES}

We have studied the effects of crossed magnetic and electric fields on skyrmion phases in a classical ferromagnetic Heisenberg model on the square lattice, including Dzyaloshinskii-Moriya interaction and a magnetoelectric coupling via the d-p hybridization mechanism. Monte Carlo simulations allowed us to map out low-temperature phase diagrams and investigate the deformation, stabilization, and suppression of skyrmion and quadrupolar textures and under both in-plane and out-of-plane electric fields. 

Our results show that the combined action of electric and magnetic fields generates a rich variety of magnetoelectric textures. In-plane electric fields tend to shrink and destabilize the skyrmion lattice, while out-of-plane fields elongate individual skyrmions and favor the emergence of bimerons. The magnetoelectric response, captured through both polarization and magnetization, reflects the intertwined symmetry and topology of the underlying spin and electric textures, offering a potential experimental pathway to identify the different chiral and multiferroic phases.

Temperature further enriches this behavior. As it increases, the system follows complex phase pathways that include intermediate bimeron-rich and skyrmion-fluid regimes. These transitions leave clear signatures in the magnetization, polarization, and chirality curves, illustrating how thermal fluctuations can partially disorder skyrmion lattices while preserving their chiral character.

A central outcome of this work is the key role played by the interplay between skyrmions and electric quadrupoles. Although these have been partially studied in \cite{Seki2012a}, here we expand the analysis to include the effects of a magnetic field in the M and P behaviours showing a similar response as in other Type II MFs \cite{Cheong2007, Cabra2025}. The close relation between these degrees of freedom governs many of the field and temperature-driven transitions. Examining the system simultaneously in the spin and dipolar spaces reveals how skyrmion deformations, bimeron formation, and phase boundaries are encoded in both magnetic and electric variables. This dual-space perspective provides a deeper understanding of magnetoelectric phenomena and underscores the importance of quadrupolar moments in mediating the coupling between spin textures and external fields.

Future studies could explore quantum effects, anisotropic interactions, and real-time dynamics under time-dependent electric fields, as well as material-specific simulations for compounds such as Cu2OSeO3 \cite{White2014, White2012, Kruchkov2017} and GaV4S8 \cite{Kezsmarki2015, Ruff2015, Bordacs2017}. Such investigations would further clarify the potential of electric-field control for low-power skyrmion-based devices.

\section{ACKNOWLEDGMENTS}

The authors thank J.S. White for discussions that motivated the present study and D. Rosales for his collaboration in the early stages of this work. This work was partially supported by CONICET (No. PIP 1146) and UNLP (No. PID X926), Argentina. F. A. G. A. is partially supported by CONICET (PIP 2021-112200200101480CO), SECyT UNLP (PI+D X947 - X1065) and Agencia I+D+i (PICT-2020-SERIE A-03205).

\bibliographystyle{apsrev4-1}

\begin{thebibliography}{83}%
\makeatletter
\providecommand \@ifxundefined [1]{%
 \@ifx{#1\undefined}
}%
\providecommand \@ifnum [1]{%
 \ifnum #1\expandafter \@firstoftwo
 \else \expandafter \@secondoftwo
 \fi
}%
\providecommand \@ifx [1]{%
 \ifx #1\expandafter \@firstoftwo
 \else \expandafter \@secondoftwo
 \fi
}%
\providecommand \natexlab [1]{#1}%
\providecommand \enquote  [1]{``#1''}%
\providecommand \bibnamefont  [1]{#1}%
\providecommand \bibfnamefont [1]{#1}%
\providecommand \citenamefont [1]{#1}%
\providecommand \href@noop [0]{\@secondoftwo}%
\providecommand \href [0]{\begingroup \@sanitize@url \@href}%
\providecommand \@href[1]{\@@startlink{#1}\@@href}%
\providecommand \@@href[1]{\endgroup#1\@@endlink}%
\providecommand \@sanitize@url [0]{\catcode `\\12\catcode `\$12\catcode `\&12\catcode `\#12\catcode `\^12\catcode `\_12\catcode `\%12\relax}%
\providecommand \@@startlink[1]{}%
\providecommand \@@endlink[0]{}%
\providecommand \url  [0]{\begingroup\@sanitize@url \@url }%
\providecommand \@url [1]{\endgroup\@href {#1}{\urlprefix }}%
\providecommand \urlprefix  [0]{URL }%
\providecommand \Eprint [0]{\href }%
\providecommand \doibase [0]{http://dx.doi.org/}%
\providecommand \selectlanguage [0]{\@gobble}%
\providecommand \bibinfo  [0]{\@secondoftwo}%
\providecommand \bibfield  [0]{\@secondoftwo}%
\providecommand \translation [1]{[#1]}%
\providecommand \BibitemOpen [0]{}%
\providecommand \bibitemStop [0]{}%
\providecommand \bibitemNoStop [0]{.\EOS\space}%
\providecommand \EOS [0]{\spacefactor3000\relax}%
\providecommand \BibitemShut  [1]{\csname bibitem#1\endcsname}%
\let\auto@bib@innerbib\@empty
\bibitem [{\citenamefont {{Bogdanov}}\ and\ \citenamefont {{Yablonskii}}(1989)}]{Bogdanov1989}%
  \BibitemOpen
  \bibfield  {author} {\bibinfo {author} {\bibfnamefont {A.~N.}\ \bibnamefont {{Bogdanov}}}\ and\ \bibinfo {author} {\bibfnamefont {D.~A.}\ \bibnamefont {{Yablonskii}}},\ }\href@noop {} {\bibfield  {journal} {\bibinfo  {journal} {Soviet Journal of Experimental and Theoretical Physics}\ }\textbf {\bibinfo {volume} {68}},\ \bibinfo {pages} {101} (\bibinfo {year} {1989})}
  \BibitemShut {NoStop}%
\bibitem [{\citenamefont {{Bogdanov}}\ and\ \citenamefont {{Hubert}}(1994)}]{Bogdanov1994}%
  \BibitemOpen
  \bibfield  {author} {\bibinfo {author} {\bibfnamefont {A.}~\bibnamefont {{Bogdanov}}}\ and\ \bibinfo {author} {\bibfnamefont {A.}~\bibnamefont {{Hubert}}},\ }\href {\doibase 10.1016/0304-8853(94)90046-9} {\bibfield  {journal} {\bibinfo  {journal} {Journal of Magnetism and Magnetic Materials}\ }\textbf {\bibinfo {volume} {138}},\ \bibinfo {pages} {255} (\bibinfo {year} {1994})}
  \BibitemShut {NoStop}%
\bibitem [{\citenamefont {Binz}\ \emph {et~al.}(2006)\citenamefont {Binz}, \citenamefont {Vishwanath},\ and\ \citenamefont {Aji}}]{Binz2006}%
  \BibitemOpen
  \bibfield  {author} {\bibinfo {author} {\bibfnamefont {B.}~\bibnamefont {Binz}}, \bibinfo {author} {\bibfnamefont {A.}~\bibnamefont {Vishwanath}}, \ and\ \bibinfo {author} {\bibfnamefont {V.}~\bibnamefont {Aji}},\ }\href {\doibase 10.1103/PhysRevLett.96.207202} {\bibfield  {journal} {\bibinfo  {journal} {Phys. Rev. Lett.}\ }\textbf {\bibinfo {volume} {96}},\ \bibinfo {pages} {207202} (\bibinfo {year} {2006})}
  \BibitemShut {NoStop}%
\bibitem [{\citenamefont {M\"uhlbauer}\ \emph {et~al.}(2009)\citenamefont {M\"uhlbauer}, \citenamefont {Binz}, \citenamefont {Jonietz}, \citenamefont {Pfleiderer}, \citenamefont {Rosch}, \citenamefont {Neubauer}, \citenamefont {Georgii},\ and\ \citenamefont {B\"oni}}]{Muhlbauer2009}%
  \BibitemOpen
  \bibfield  {author} {\bibinfo {author} {\bibfnamefont {S.}~\bibnamefont {M\"uhlbauer}}, \bibinfo {author} {\bibfnamefont {B.}~\bibnamefont {Binz}}, \bibinfo {author} {\bibfnamefont {F.}~\bibnamefont {Jonietz}}, \bibinfo {author} {\bibfnamefont {C.}~\bibnamefont {Pfleiderer}}, \bibinfo {author} {\bibfnamefont {A.}~\bibnamefont {Rosch}}, \bibinfo {author} {\bibfnamefont {A.}~\bibnamefont {Neubauer}}, \bibinfo {author} {\bibfnamefont {R.}~\bibnamefont {Georgii}}, \ and\ \bibinfo {author} {\bibfnamefont {P.}~\bibnamefont {B\"oni}},\ }\href {\doibase 10.1126/science.1166767} {\bibfield  {journal} {\bibinfo  {journal} {Science}\ }\textbf {\bibinfo {volume} {323}},\ \bibinfo {pages} {915} (\bibinfo {year} {2009})}
  \BibitemShut {NoStop}%
\bibitem [{\citenamefont {Neubauer}\ \emph {et~al.}(2009)\citenamefont {Neubauer}, \citenamefont {Pfleiderer}, \citenamefont {Binz}, \citenamefont {Rosch}, \citenamefont {Ritz}, \citenamefont {Niklowitz},\ and\ \citenamefont {B\"oni}}]{Neubauer2009}%
  \BibitemOpen
  \bibfield  {author} {\bibinfo {author} {\bibfnamefont {A.}~\bibnamefont {Neubauer}}, \bibinfo {author} {\bibfnamefont {C.}~\bibnamefont {Pfleiderer}}, \bibinfo {author} {\bibfnamefont {B.}~\bibnamefont {Binz}}, \bibinfo {author} {\bibfnamefont {A.}~\bibnamefont {Rosch}}, \bibinfo {author} {\bibfnamefont {R.}~\bibnamefont {Ritz}}, \bibinfo {author} {\bibfnamefont {P.~G.}\ \bibnamefont {Niklowitz}}, \ and\ \bibinfo {author} {\bibfnamefont {P.}~\bibnamefont {B\"oni}},\ }\href {\doibase 10.1103/PhysRevLett.102.186602} {\bibfield  {journal} {\bibinfo  {journal} {Phys. Rev. Lett.}\ }\textbf {\bibinfo {volume} {102}},\ \bibinfo {pages} {186602} (\bibinfo {year} {2009})}
  \BibitemShut {NoStop}%
\bibitem [{\citenamefont {M\"unzer}\ \emph {et~al.}(2010)\citenamefont {M\"unzer}, \citenamefont {Neubauer}, \citenamefont {Adams}, \citenamefont {M\"uhlbauer}, \citenamefont {Franz}, \citenamefont {Jonietz}, \citenamefont {Georgii}, \citenamefont {B\"oni}, \citenamefont {Pedersen}, \citenamefont {Schmidt}, \citenamefont {Rosch},\ and\ \citenamefont {Pfleiderer}}]{Munzer2010}%
  \BibitemOpen
  \bibfield  {author} {\bibinfo {author} {\bibfnamefont {W.}~\bibnamefont {M\"unzer}}, \bibinfo {author} {\bibfnamefont {A.}~\bibnamefont {Neubauer}}, \bibinfo {author} {\bibfnamefont {T.}~\bibnamefont {Adams}}, \bibinfo {author} {\bibfnamefont {S.}~\bibnamefont {M\"uhlbauer}}, \bibinfo {author} {\bibfnamefont {C.}~\bibnamefont {Franz}}, \bibinfo {author} {\bibfnamefont {F.}~\bibnamefont {Jonietz}}, \bibinfo {author} {\bibfnamefont {R.}~\bibnamefont {Georgii}}, \bibinfo {author} {\bibfnamefont {P.}~\bibnamefont {B\"oni}}, \bibinfo {author} {\bibfnamefont {B.}~\bibnamefont {Pedersen}}, \bibinfo {author} {\bibfnamefont {M.}~\bibnamefont {Schmidt}}, \bibinfo {author} {\bibfnamefont {A.}~\bibnamefont {Rosch}}, \ and\ \bibinfo {author} {\bibfnamefont {C.}~\bibnamefont {Pfleiderer}},\ }\href {\doibase 10.1103/PhysRevB.81.041203} {\bibfield  {journal} {\bibinfo  {journal} {Phys. Rev. B}\ }\textbf {\bibinfo {volume} {81}},\ \bibinfo {pages} {041203} (\bibinfo {year} {2010})}
  \BibitemShut {NoStop}%
\bibitem [{\citenamefont {Yu}\ \emph {et~al.}(2010)\citenamefont {Yu}, \citenamefont {Onose}, \citenamefont {Kanazawa}, \citenamefont {Park}, \citenamefont {Han}, \citenamefont {Matsui}, \citenamefont {Nagaosa},\ and\ \citenamefont {Tokura}}]{Yu2010}%
  \BibitemOpen
  \bibfield  {author} {\bibinfo {author} {\bibfnamefont {X.~Z.}\ \bibnamefont {Yu}}, \bibinfo {author} {\bibfnamefont {Y.}~\bibnamefont {Onose}}, \bibinfo {author} {\bibfnamefont {N.}~\bibnamefont {Kanazawa}}, \bibinfo {author} {\bibfnamefont {J.-H.}\ \bibnamefont {Park}}, \bibinfo {author} {\bibfnamefont {J.~H.}\ \bibnamefont {Han}}, \bibinfo {author} {\bibfnamefont {Y.}~\bibnamefont {Matsui}}, \bibinfo {author} {\bibfnamefont {N.}~\bibnamefont {Nagaosa}}, \ and\ \bibinfo {author} {\bibfnamefont {Y.}~\bibnamefont {Tokura}},\ }\href {https://api.semanticscholar.org/CorpusID:205220974} {\bibfield  {journal} {\bibinfo  {journal} {Nature}\ }\textbf {\bibinfo {volume} {465}},\ \bibinfo {pages} {901} (\bibinfo {year} {2010})}
  \BibitemShut {NoStop}%
\bibitem [{\citenamefont {Yu}\ \emph {et~al.}(2011)\citenamefont {Yu}, \citenamefont {Kanazawa}, \citenamefont {Onose}, \citenamefont {Kimoto}, \citenamefont {Zhang}, \citenamefont {Ishiwata}, \citenamefont {Matsui},\ and\ \citenamefont {Tokura}}]{Yu2011}%
  \BibitemOpen
  \bibfield  {author} {\bibinfo {author} {\bibfnamefont {X.}~\bibnamefont {Yu}}, \bibinfo {author} {\bibfnamefont {N.}~\bibnamefont {Kanazawa}}, \bibinfo {author} {\bibfnamefont {Y.}~\bibnamefont {Onose}}, \bibinfo {author} {\bibfnamefont {K.}~\bibnamefont {Kimoto}}, \bibinfo {author} {\bibfnamefont {W.}~\bibnamefont {Zhang}}, \bibinfo {author} {\bibfnamefont {S.}~\bibnamefont {Ishiwata}}, \bibinfo {author} {\bibfnamefont {Y.}~\bibnamefont {Matsui}}, \ and\ \bibinfo {author} {\bibfnamefont {Y.}~\bibnamefont {Tokura}},\ }\href {https://api.semanticscholar.org/CorpusID:19433416} {\bibfield  {journal} {\bibinfo  {journal} {Nature materials}\ }\textbf {\bibinfo {volume} {10 2}},\ \bibinfo {pages} {106} (\bibinfo {year} {2011})}
  \BibitemShut {NoStop}%
\bibitem [{\citenamefont {Yu}\ \emph {et~al.}(2012)\citenamefont {Yu}, \citenamefont {Mostovoy}, \citenamefont {Tokunaga}, \citenamefont {Zhang}, \citenamefont {Kimoto}, \citenamefont {Matsui}, \citenamefont {Kaneko}, \citenamefont {Nagaosa},\ and\ \citenamefont {Tokura}}]{Yu2012}%
  \BibitemOpen
  \bibfield  {author} {\bibinfo {author} {\bibfnamefont {X.}~\bibnamefont {Yu}}, \bibinfo {author} {\bibfnamefont {M.}~\bibnamefont {Mostovoy}}, \bibinfo {author} {\bibfnamefont {Y.}~\bibnamefont {Tokunaga}}, \bibinfo {author} {\bibfnamefont {W.}~\bibnamefont {Zhang}}, \bibinfo {author} {\bibfnamefont {K.}~\bibnamefont {Kimoto}}, \bibinfo {author} {\bibfnamefont {Y.}~\bibnamefont {Matsui}}, \bibinfo {author} {\bibfnamefont {Y.}~\bibnamefont {Kaneko}}, \bibinfo {author} {\bibfnamefont {N.}~\bibnamefont {Nagaosa}}, \ and\ \bibinfo {author} {\bibfnamefont {Y.}~\bibnamefont {Tokura}},\ }\href {https://api.semanticscholar.org/CorpusID:4998742} {\bibfield  {journal} {\bibinfo  {journal} {Proceedings of the National Academy of Sciences}\ }\textbf {\bibinfo {volume} {109}},\ \bibinfo {pages} {8856 } (\bibinfo {year} {2012})}
  \BibitemShut {NoStop}%
\bibitem [{\citenamefont {Woo}\ \emph {et~al.}(2016)\citenamefont {Woo}, \citenamefont {Litzius}, \citenamefont {Krueger}, \citenamefont {Im}, \citenamefont {Caretta}, \citenamefont {Richter}, \citenamefont {Mann}, \citenamefont {Krone}, \citenamefont {Reeve}, \citenamefont {Weigand}, \citenamefont {Agrawal}, \citenamefont {Lemesh}, \citenamefont {Mawass}, \citenamefont {Fischer}, \citenamefont {Kl{\"a}ui},\ and\ \citenamefont {Beach}}]{Woo2016}%
  \BibitemOpen
  \bibfield  {author} {\bibinfo {author} {\bibfnamefont {S.}~\bibnamefont {Woo}}, \bibinfo {author} {\bibfnamefont {K.}~\bibnamefont {Litzius}}, \bibinfo {author} {\bibfnamefont {B.}~\bibnamefont {Krueger}}, \bibinfo {author} {\bibfnamefont {M.-Y.}\ \bibnamefont {Im}}, \bibinfo {author} {\bibfnamefont {L.}~\bibnamefont {Caretta}}, \bibinfo {author} {\bibfnamefont {K.}~\bibnamefont {Richter}}, \bibinfo {author} {\bibfnamefont {M.}~\bibnamefont {Mann}}, \bibinfo {author} {\bibfnamefont {A.}~\bibnamefont {Krone}}, \bibinfo {author} {\bibfnamefont {R.~M.}\ \bibnamefont {Reeve}}, \bibinfo {author} {\bibfnamefont {M.}~\bibnamefont {Weigand}}, \bibinfo {author} {\bibfnamefont {P.}~\bibnamefont {Agrawal}}, \bibinfo {author} {\bibfnamefont {I.}~\bibnamefont {Lemesh}}, \bibinfo {author} {\bibfnamefont {M.-A.}\ \bibnamefont {Mawass}}, \bibinfo {author} {\bibfnamefont {P.}~\bibnamefont {Fischer}}, \bibinfo {author} {\bibfnamefont {M.}~\bibnamefont {Kl{\"a}ui}}, \ and\ \bibinfo {author} {\bibfnamefont {G.~S.~D.}\
  \bibnamefont {Beach}},\ }\href {https://api.semanticscholar.org/CorpusID:205413036} {\bibfield  {journal} {\bibinfo  {journal} {Nature materials}\ }\textbf {\bibinfo {volume} {15 5}},\ \bibinfo {pages} {501} (\bibinfo {year} {2016})}
  \BibitemShut {NoStop}%
\bibitem [{\citenamefont {Karube}\ \emph {et~al.}(2018)\citenamefont {Karube}, \citenamefont {White}, \citenamefont {Morikawa}, \citenamefont {Dewhurst}, \citenamefont {Cubitt}, \citenamefont {Kikkawa}, \citenamefont {Yu}, \citenamefont {Tokunaga}, \citenamefont {hisa Arima}, \citenamefont {R\o{}nnow}, \citenamefont {Tokura},\ and\ \citenamefont {Taguchi}}]{Karube2018}%
  \BibitemOpen
  \bibfield  {author} {\bibinfo {author} {\bibfnamefont {K.}~\bibnamefont {Karube}}, \bibinfo {author} {\bibfnamefont {J.~S.}\ \bibnamefont {White}}, \bibinfo {author} {\bibfnamefont {D.}~\bibnamefont {Morikawa}}, \bibinfo {author} {\bibfnamefont {C.~D.}\ \bibnamefont {Dewhurst}}, \bibinfo {author} {\bibfnamefont {R.}~\bibnamefont {Cubitt}}, \bibinfo {author} {\bibfnamefont {A.}~\bibnamefont {Kikkawa}}, \bibinfo {author} {\bibfnamefont {X.}~\bibnamefont {Yu}}, \bibinfo {author} {\bibfnamefont {Y.}~\bibnamefont {Tokunaga}}, \bibinfo {author} {\bibfnamefont {T.}~\bibnamefont {hisa Arima}}, \bibinfo {author} {\bibfnamefont {H.~M.}\ \bibnamefont {R\o{}nnow}}, \bibinfo {author} {\bibfnamefont {Y.}~\bibnamefont {Tokura}}, \ and\ \bibinfo {author} {\bibfnamefont {Y.}~\bibnamefont {Taguchi}},\ }\href {\doibase 10.1126/sciadv.aar7043} {\bibfield  {journal} {\bibinfo  {journal} {Science Advances}\ }\textbf {\bibinfo {volume} {4}},\ \bibinfo {pages} {eaar7043} (\bibinfo {year} {2018})}
  \BibitemShut {NoStop}%
\bibitem [{\citenamefont {Nagaosa}\ and\ \citenamefont {Tokura}(2013)}]{Nagaosa2013}%
  \BibitemOpen
  \bibfield  {author} {\bibinfo {author} {\bibfnamefont {N.}~\bibnamefont {Nagaosa}}\ and\ \bibinfo {author} {\bibfnamefont {Y.}~\bibnamefont {Tokura}},\ }\href {https://api.semanticscholar.org/CorpusID:3073721} {\bibfield  {journal} {\bibinfo  {journal} {Nature nanotechnology}\ }\textbf {\bibinfo {volume} {8 12}},\ \bibinfo {pages} {899} (\bibinfo {year} {2013})}
  \BibitemShut {NoStop}%
\bibitem [{\citenamefont {Fert}\ \emph {et~al.}(2013)\citenamefont {Fert}, \citenamefont {Cros},\ and\ \citenamefont {Sampaio}}]{Fert2013}%
  \BibitemOpen
  \bibfield  {author} {\bibinfo {author} {\bibfnamefont {A.}~\bibnamefont {Fert}}, \bibinfo {author} {\bibfnamefont {V.}~\bibnamefont {Cros}}, \ and\ \bibinfo {author} {\bibfnamefont {J.}~\bibnamefont {Sampaio}},\ }\href {https://api.semanticscholar.org/CorpusID:42822349} {\bibfield  {journal} {\bibinfo  {journal} {Nature nanotechnology}\ }\textbf {\bibinfo {volume} {8 3}},\ \bibinfo {pages} {152} (\bibinfo {year} {2013})}
  \BibitemShut {NoStop}%
\bibitem [{\citenamefont {Back}\ \emph {et~al.}(2020)\citenamefont {Back}, \citenamefont {Cros}, \citenamefont {Ebert}, \citenamefont {Everschor-Sitte}, \citenamefont {Fert}, \citenamefont {Garst}, \citenamefont {Ma}, \citenamefont {Mankovsky}, \citenamefont {Monchesky}, \citenamefont {Mostovoy}, \citenamefont {Nagaosa}, \citenamefont {Parkin}, \citenamefont {Pfleiderer}, \citenamefont {Reyren}, \citenamefont {Rosch}, \citenamefont {Taguchi}, \citenamefont {Tokura}, \citenamefont {von Bergmann},\ and\ \citenamefont {Zang}}]{Back2020}%
  \BibitemOpen
  \bibfield  {author} {\bibinfo {author} {\bibfnamefont {C.}~\bibnamefont {Back}}, \bibinfo {author} {\bibfnamefont {V.}~\bibnamefont {Cros}}, \bibinfo {author} {\bibfnamefont {H.}~\bibnamefont {Ebert}}, \bibinfo {author} {\bibfnamefont {K.}~\bibnamefont {Everschor-Sitte}}, \bibinfo {author} {\bibfnamefont {A.}~\bibnamefont {Fert}}, \bibinfo {author} {\bibfnamefont {M.}~\bibnamefont {Garst}}, \bibinfo {author} {\bibfnamefont {T.}~\bibnamefont {Ma}}, \bibinfo {author} {\bibfnamefont {S.}~\bibnamefont {Mankovsky}}, \bibinfo {author} {\bibfnamefont {T.~L.}\ \bibnamefont {Monchesky}}, \bibinfo {author} {\bibfnamefont {M.}~\bibnamefont {Mostovoy}}, \bibinfo {author} {\bibfnamefont {N.}~\bibnamefont {Nagaosa}}, \bibinfo {author} {\bibfnamefont {S.~S.~P.}\ \bibnamefont {Parkin}}, \bibinfo {author} {\bibfnamefont {C.}~\bibnamefont {Pfleiderer}}, \bibinfo {author} {\bibfnamefont {N.}~\bibnamefont {Reyren}}, \bibinfo {author} {\bibfnamefont {A.}~\bibnamefont {Rosch}}, \bibinfo {author} {\bibfnamefont {Y.}~\bibnamefont
  {Taguchi}}, \bibinfo {author} {\bibfnamefont {Y.}~\bibnamefont {Tokura}}, \bibinfo {author} {\bibfnamefont {K.}~\bibnamefont {von Bergmann}}, \ and\ \bibinfo {author} {\bibfnamefont {J.}~\bibnamefont {Zang}},\ }\href {\doibase 10.1088/1361-6463/ab8418} {\bibfield  {journal} {\bibinfo  {journal} {Journal of Physics D: Applied Physics}\ }\textbf {\bibinfo {volume} {53}},\ \bibinfo {pages} {363001} (\bibinfo {year} {2020})}
  \BibitemShut {NoStop}%
\bibitem [{\citenamefont {GÃ¶bel}\ \emph {et~al.}(2021)\citenamefont {GÃ¶bel}, \citenamefont {Mertig},\ and\ \citenamefont {Tretiakov}}]{Gobel2021}%
  \BibitemOpen
  \bibfield  {author} {\bibinfo {author} {\bibfnamefont {B.}~\bibnamefont {G\"obel}}, \bibinfo {author} {\bibfnamefont {I.}~\bibnamefont {Mertig}}, \ and\ \bibinfo {author} {\bibfnamefont {O.~A.}\ \bibnamefont {Tretiakov}},\ }\href {\doibase https://doi.org/10.1016/j.physrep.2020.10.001} {\bibfield  {journal} {\bibinfo  {journal} {Physics Reports}\ }\textbf {\bibinfo {volume} {895}},\ \bibinfo {pages} {1} (\bibinfo {year} {2021})}
  \BibitemShut {NoStop}%
\bibitem [{\citenamefont {Psaroudaki}\ and\ \citenamefont {Panagopoulos}(2021)}]{Psaroudaki2021}%
  \BibitemOpen
  \bibfield  {author} {\bibinfo {author} {\bibfnamefont {C.}~\bibnamefont {Psaroudaki}}\ and\ \bibinfo {author} {\bibfnamefont {C.}~\bibnamefont {Panagopoulos}},\ }\href {\doibase 10.1103/PhysRevLett.127.067201} {\bibfield  {journal} {\bibinfo  {journal} {Phys. Rev. Lett.}\ }\textbf {\bibinfo {volume} {127}},\ \bibinfo {pages} {067201} (\bibinfo {year} {2021})}
  \BibitemShut {NoStop}%
\bibitem [{\citenamefont {Psaroudaki}\ \emph {et~al.}(2023)\citenamefont {Psaroudaki}, \citenamefont {Peraticos},\ and\ \citenamefont {Panagopoulos}}]{Psaroudaki2023}%
  \BibitemOpen
  \bibfield  {author} {\bibinfo {author} {\bibfnamefont {C.}~\bibnamefont {Psaroudaki}}, \bibinfo {author} {\bibfnamefont {E.}~\bibnamefont {Peraticos}}, \ and\ \bibinfo {author} {\bibfnamefont {C.}~\bibnamefont {Panagopoulos}},\ }\href@noop {} {\bibfield  {journal} {\bibinfo  {journal} {Applied Physics Letters}\ }\textbf {\bibinfo {volume} {123}} (\bibinfo {year} {2023})}
  \BibitemShut {NoStop}%
\bibitem [{\citenamefont {Xia}\ \emph {et~al.}(2023)\citenamefont {Xia}, \citenamefont {Zhang}, \citenamefont {Liu}, \citenamefont {Zhou},\ and\ \citenamefont {Ezawa}}]{Xia2023}%
  \BibitemOpen
  \bibfield  {author} {\bibinfo {author} {\bibfnamefont {J.}~\bibnamefont {Xia}}, \bibinfo {author} {\bibfnamefont {X.}~\bibnamefont {Zhang}}, \bibinfo {author} {\bibfnamefont {X.}~\bibnamefont {Liu}}, \bibinfo {author} {\bibfnamefont {Y.}~\bibnamefont {Zhou}}, \ and\ \bibinfo {author} {\bibfnamefont {M.}~\bibnamefont {Ezawa}},\ }\href {\doibase 10.1103/PhysRevLett.130.106701} {\bibfield  {journal} {\bibinfo  {journal} {Phys. Rev. Lett.}\ }\textbf {\bibinfo {volume} {130}},\ \bibinfo {pages} {106701} (\bibinfo {year} {2023})}
  \BibitemShut {NoStop}%
\bibitem [{\citenamefont {Pappas}\ \emph {et~al.}(2009)\citenamefont {Pappas}, \citenamefont {Leli\`evre-Berna}, \citenamefont {Falus}, \citenamefont {Bentley}, \citenamefont {Moskvin}, \citenamefont {Grigoriev}, \citenamefont {Fouquet},\ and\ \citenamefont {Farago}}]{Pappas2009}%
  \BibitemOpen
  \bibfield  {author} {\bibinfo {author} {\bibfnamefont {C.}~\bibnamefont {Pappas}}, \bibinfo {author} {\bibfnamefont {E.}~\bibnamefont {Leli\`evre-Berna}}, \bibinfo {author} {\bibfnamefont {P.}~\bibnamefont {Falus}}, \bibinfo {author} {\bibfnamefont {P.~M.}\ \bibnamefont {Bentley}}, \bibinfo {author} {\bibfnamefont {E.}~\bibnamefont {Moskvin}}, \bibinfo {author} {\bibfnamefont {S.}~\bibnamefont {Grigoriev}}, \bibinfo {author} {\bibfnamefont {P.}~\bibnamefont {Fouquet}}, \ and\ \bibinfo {author} {\bibfnamefont {B.}~\bibnamefont {Farago}},\ }\href {\doibase 10.1103/PhysRevLett.102.197202} {\bibfield  {journal} {\bibinfo  {journal} {Phys. Rev. Lett.}\ }\textbf {\bibinfo {volume} {102}},\ \bibinfo {pages} {197202} (\bibinfo {year} {2009})}
  \BibitemShut {NoStop}%
\bibitem [{\citenamefont {Banerjee}\ \emph {et~al.}(2014)\citenamefont {Banerjee}, \citenamefont {Rowland}, \citenamefont {Erten},\ and\ \citenamefont {Randeria}}]{Benerjee2014}%
  \BibitemOpen
  \bibfield  {author} {\bibinfo {author} {\bibfnamefont {S.}~\bibnamefont {Banerjee}}, \bibinfo {author} {\bibfnamefont {J.}~\bibnamefont {Rowland}}, \bibinfo {author} {\bibfnamefont {O.}~\bibnamefont {Erten}}, \ and\ \bibinfo {author} {\bibfnamefont {M.}~\bibnamefont {Randeria}},\ }\href {\doibase 10.1103/PhysRevX.4.031045} {\bibfield  {journal} {\bibinfo  {journal} {Phys. Rev. X}\ }\textbf {\bibinfo {volume} {4}},\ \bibinfo {pages} {031045} (\bibinfo {year} {2014})}
  \BibitemShut {NoStop}%
\bibitem [{\citenamefont {Yi}\ \emph {et~al.}(2009)\citenamefont {Yi}, \citenamefont {Onoda}, \citenamefont {Nagaosa},\ and\ \citenamefont {Han}}]{Yi2009}%
  \BibitemOpen
  \bibfield  {author} {\bibinfo {author} {\bibfnamefont {S.~D.}\ \bibnamefont {Yi}}, \bibinfo {author} {\bibfnamefont {S.}~\bibnamefont {Onoda}}, \bibinfo {author} {\bibfnamefont {N.}~\bibnamefont {Nagaosa}}, \ and\ \bibinfo {author} {\bibfnamefont {J.~H.}\ \bibnamefont {Han}},\ }\href {\doibase 10.1103/PhysRevB.80.054416} {\bibfield  {journal} {\bibinfo  {journal} {Phys. Rev. B}\ }\textbf {\bibinfo {volume} {80}},\ \bibinfo {pages} {054416} (\bibinfo {year} {2009})}
  \BibitemShut {NoStop}%
\bibitem [{\citenamefont {Buhrandt}\ and\ \citenamefont {Fritz}(2013)}]{Buhrandt2013}%
  \BibitemOpen
  \bibfield  {author} {\bibinfo {author} {\bibfnamefont {S.}~\bibnamefont {Buhrandt}}\ and\ \bibinfo {author} {\bibfnamefont {L.}~\bibnamefont {Fritz}},\ }\href {\doibase 10.1103/PhysRevB.88.195137} {\bibfield  {journal} {\bibinfo  {journal} {Phys. Rev. B}\ }\textbf {\bibinfo {volume} {88}},\ \bibinfo {pages} {195137} (\bibinfo {year} {2013})}
  \BibitemShut {NoStop}%
\bibitem [{\citenamefont {Huang}\ and\ \citenamefont {Chien}(2012)}]{Huang2012}%
  \BibitemOpen
  \bibfield  {author} {\bibinfo {author} {\bibfnamefont {S.~X.}\ \bibnamefont {Huang}}\ and\ \bibinfo {author} {\bibfnamefont {C.~L.}\ \bibnamefont {Chien}},\ }\href {\doibase 10.1103/PhysRevLett.108.267201} {\bibfield  {journal} {\bibinfo  {journal} {Phys. Rev. Lett.}\ }\textbf {\bibinfo {volume} {108}},\ \bibinfo {pages} {267201} (\bibinfo {year} {2012})}
  \BibitemShut {NoStop}%
\bibitem [{\citenamefont {Romming}\ \emph {et~al.}(2013)\citenamefont {Romming}, \citenamefont {Hanneken}, \citenamefont {Menzel}, \citenamefont {Bickel}, \citenamefont {Wolter}, \citenamefont {von Bergmann}, \citenamefont {Kubetzka},\ and\ \citenamefont {Wiesendanger}}]{Romming2013}%
  \BibitemOpen
  \bibfield  {author} {\bibinfo {author} {\bibfnamefont {N.}~\bibnamefont {Romming}}, \bibinfo {author} {\bibfnamefont {C.}~\bibnamefont {Hanneken}}, \bibinfo {author} {\bibfnamefont {M.}~\bibnamefont {Menzel}}, \bibinfo {author} {\bibfnamefont {J.~E.}\ \bibnamefont {Bickel}}, \bibinfo {author} {\bibfnamefont {B.}~\bibnamefont {Wolter}}, \bibinfo {author} {\bibfnamefont {K.}~\bibnamefont {von Bergmann}}, \bibinfo {author} {\bibfnamefont {A.}~\bibnamefont {Kubetzka}}, \ and\ \bibinfo {author} {\bibfnamefont {R.}~\bibnamefont {Wiesendanger}},\ }\href {\doibase 10.1126/science.1240573} {\bibfield  {journal} {\bibinfo  {journal} {Science}\ }\textbf {\bibinfo {volume} {341}},\ \bibinfo {pages} {636} (\bibinfo {year} {2013})}
  \BibitemShut {NoStop}%
\bibitem [{\citenamefont {Romming}\ \emph {et~al.}(2015)\citenamefont {Romming}, \citenamefont {Kubetzka}, \citenamefont {Hanneken}, \citenamefont {von Bergmann},\ and\ \citenamefont {Wiesendanger}}]{Romming2015}%
  \BibitemOpen
  \bibfield  {author} {\bibinfo {author} {\bibfnamefont {N.}~\bibnamefont {Romming}}, \bibinfo {author} {\bibfnamefont {A.}~\bibnamefont {Kubetzka}}, \bibinfo {author} {\bibfnamefont {C.}~\bibnamefont {Hanneken}}, \bibinfo {author} {\bibfnamefont {K.}~\bibnamefont {von Bergmann}}, \ and\ \bibinfo {author} {\bibfnamefont {R.}~\bibnamefont {Wiesendanger}},\ }\href {\doibase 10.1103/PhysRevLett.114.177203} {\bibfield  {journal} {\bibinfo  {journal} {Phys. Rev. Lett.}\ }\textbf {\bibinfo {volume} {114}},\ \bibinfo {pages} {177203} (\bibinfo {year} {2015})}
  \BibitemShut {NoStop}%
\bibitem [{\citenamefont {Iwasaki}\ \emph {et~al.}(2013)\citenamefont {Iwasaki}, \citenamefont {Mochizuki},\ and\ \citenamefont {Nagaosa}}]{Iwasaki2013}%
  \BibitemOpen
  \bibfield  {author} {\bibinfo {author} {\bibfnamefont {J.}~\bibnamefont {Iwasaki}}, \bibinfo {author} {\bibfnamefont {M.}~\bibnamefont {Mochizuki}}, \ and\ \bibinfo {author} {\bibfnamefont {N.}~\bibnamefont {Nagaosa}},\ }\href {https://api.semanticscholar.org/CorpusID:5207271} {\bibfield  {journal} {\bibinfo  {journal} {Nature Communications}\ }\textbf {\bibinfo {volume} {4}} (\bibinfo {year} {2013})}
  \BibitemShut {NoStop}%
\bibitem [{\citenamefont {Hayami}(2022)}]{Hayami2022}%
  \BibitemOpen
  \bibfield  {author} {\bibinfo {author} {\bibfnamefont {S.}~\bibnamefont {Hayami}},\ }\href {\doibase 10.1103/PhysRevB.105.014408} {\bibfield  {journal} {\bibinfo  {journal} {Phys. Rev. B}\ }\textbf {\bibinfo {volume} {105}},\ \bibinfo {pages} {014408} (\bibinfo {year} {2022})}
  \BibitemShut {NoStop}%
\bibitem [{\citenamefont {Heinze}\ \emph {et~al.}(2011)\citenamefont {Heinze}, \citenamefont {von Bergmann}, \citenamefont {Menzel}, \citenamefont {Brede}, \citenamefont {Kubetzka}, \citenamefont {Wiesendanger}, \citenamefont {Bihlmayer},\ and\ \citenamefont {Bl{\"u}gel}}]{Heinze2011}%
  \BibitemOpen
  \bibfield  {author} {\bibinfo {author} {\bibfnamefont {S.}~\bibnamefont {Heinze}}, \bibinfo {author} {\bibfnamefont {K.}~\bibnamefont {von Bergmann}}, \bibinfo {author} {\bibfnamefont {M.}~\bibnamefont {Menzel}}, \bibinfo {author} {\bibfnamefont {J.}~\bibnamefont {Brede}}, \bibinfo {author} {\bibfnamefont {A.}~\bibnamefont {Kubetzka}}, \bibinfo {author} {\bibfnamefont {R.}~\bibnamefont {Wiesendanger}}, \bibinfo {author} {\bibfnamefont {G.}~\bibnamefont {Bihlmayer}}, \ and\ \bibinfo {author} {\bibfnamefont {S.}~\bibnamefont {Bl{\"u}gel}},\ }\href {https://api.semanticscholar.org/CorpusID:123676430} {\bibfield  {journal} {\bibinfo  {journal} {Nature Physics}\ }\textbf {\bibinfo {volume} {7}},\ \bibinfo {pages} {713} (\bibinfo {year} {2011})}
  \BibitemShut {NoStop}%
\bibitem [{\citenamefont {Dzyaloshinsky}(1958)}]{Dzyaloshinskii1958}%
  \BibitemOpen
  \bibfield  {author} {\bibinfo {author} {\bibfnamefont {I.}~\bibnamefont {Dzyaloshinsky}},\ }\href {\doibase https://doi.org/10.1016/0022-3697(58)90076-3} {\bibfield  {journal} {\bibinfo  {journal} {Journal of Physics and Chemistry of Solids}\ }\textbf {\bibinfo {volume} {4}},\ \bibinfo {pages} {241} (\bibinfo {year} {1958})}
  \BibitemShut {NoStop}%
\bibitem [{\citenamefont {Moriya}(1960)}]{Moriya1960}%
  \BibitemOpen
  \bibfield  {author} {\bibinfo {author} {\bibfnamefont {T.}~\bibnamefont {Moriya}},\ }\href {\doibase 10.1103/PhysRev.120.91} {\bibfield  {journal} {\bibinfo  {journal} {Phys. Rev.}\ }\textbf {\bibinfo {volume} {120}},\ \bibinfo {pages} {91} (\bibinfo {year} {1960})}
  \BibitemShut {NoStop}%
\bibitem [{\citenamefont {Okubo}\ \emph {et~al.}(2012)\citenamefont {Okubo}, \citenamefont {Chung},\ and\ \citenamefont {Kawamura}}]{Okubo2012}%
  \BibitemOpen
  \bibfield  {author} {\bibinfo {author} {\bibfnamefont {T.}~\bibnamefont {Okubo}}, \bibinfo {author} {\bibfnamefont {S.}~\bibnamefont {Chung}}, \ and\ \bibinfo {author} {\bibfnamefont {H.}~\bibnamefont {Kawamura}},\ }\href {https://link.aps.org/doi/10.1103/PhysRevLett.108.017206} {\bibfield  {journal} {\bibinfo  {journal} {Phys. Rev. Lett.}\ }\textbf {\bibinfo {volume} {108}},\ \bibinfo {pages} {017206} (\bibinfo {year} {2012})}
  \BibitemShut {NoStop}%
\bibitem [{\citenamefont {Leonov}(2015)}]{Leonov2015}%
  \BibitemOpen
  \bibfield  {author} {\bibinfo {author} {\bibfnamefont {A.}~\bibnamefont {Leonov}},\ }\href {\doibase 10.1038/ncomms9275} {\bibfield  {journal} {\bibinfo  {journal} {Nature Communications}\ }\textbf {\bibinfo {volume} {6}},\ \bibinfo {pages} {8275} (\bibinfo {year} {2015})}
  \BibitemShut {NoStop}%
\bibitem [{\citenamefont {Mohylna}\ \emph {et~al.}(2022)\citenamefont {Mohylna}, \citenamefont {G{\'o}mez~Albarrac{\'\i}n}, \citenamefont {{\v{Z}}ukovi{\v{c}}},\ and\ \citenamefont {Rosales}}]{Mohylna2022}%
  \BibitemOpen
  \bibfield  {author} {\bibinfo {author} {\bibfnamefont {M.}~\bibnamefont {Mohylna}}, \bibinfo {author} {\bibfnamefont {F.~A.}\ \bibnamefont {G{\'o}mez~Albarrac{\'\i}n}}, \bibinfo {author} {\bibfnamefont {M.}~\bibnamefont {{\v{Z}}ukovi{\v{c}}}}, \ and\ \bibinfo {author} {\bibfnamefont {H.~D.}\ \bibnamefont {Rosales}},\ }\href@noop {} {\bibfield  {journal} {\bibinfo  {journal} {Physical Review B}\ }\textbf {\bibinfo {volume} {106}},\ \bibinfo {pages} {224406} (\bibinfo {year} {2022})}
  \BibitemShut {NoStop}%
\bibitem [{\citenamefont {Gao}\ \emph {et~al.}(2020)\citenamefont {Gao}, \citenamefont {Rosales}, \citenamefont {G{\'o}mez~Albarrac{\'\i}n}, \citenamefont {Tsurkan}, \citenamefont {Kaur}, \citenamefont {Fennell}, \citenamefont {Steffens}, \citenamefont {Boehm}, \citenamefont {{\v{C}}erm{\'a}k}, \citenamefont {Schneidewind}, \citenamefont {Ressouche}, \citenamefont {Cabra}, \citenamefont {R\"uegg},\ and\ \citenamefont {Zakarko}}]{Gao2020}%
  \BibitemOpen
  \bibfield  {author} {\bibinfo {author} {\bibfnamefont {S.}~\bibnamefont {Gao}}, \bibinfo {author} {\bibfnamefont {H.~D.}\ \bibnamefont {Rosales}}, \bibinfo {author} {\bibfnamefont {F.~A.}\ \bibnamefont {G{\'o}mez~Albarrac{\'\i}n}}, \bibinfo {author} {\bibfnamefont {V.}~\bibnamefont {Tsurkan}}, \bibinfo {author} {\bibfnamefont {G.}~\bibnamefont {Kaur}}, \bibinfo {author} {\bibfnamefont {T.}~\bibnamefont {Fennell}}, \bibinfo {author} {\bibfnamefont {P.}~\bibnamefont {Steffens}}, \bibinfo {author} {\bibfnamefont {M.}~\bibnamefont {Boehm}}, \bibinfo {author} {\bibfnamefont {P.}~\bibnamefont {{\v{C}}erm{\'a}k}}, \bibinfo {author} {\bibfnamefont {A.}~\bibnamefont {Schneidewind}}, \bibinfo {author} {\bibfnamefont {E.}~\bibnamefont {Ressouche}}, \bibinfo {author} {\bibfnamefont {D.~C.}\ \bibnamefont {Cabra}}, \bibinfo {author} {\bibfnamefont {C.}~\bibnamefont {R\"uegg}}, \ and\ \bibinfo {author} {\bibfnamefont {O.}~\bibnamefont {Zakarko}},\ }\href {https://doi.org/10.1038/s41586-020-2716-8} {\bibfield  {journal}
  {\bibinfo  {journal} {Nature}\ }\textbf {\bibinfo {volume} {586}},\ \bibinfo {pages} {37} (\bibinfo {year} {2020})}
  \BibitemShut {NoStop}%
\bibitem [{\citenamefont {Rosales}\ \emph {et~al.}(2022)\citenamefont {Rosales}, \citenamefont {Albarrac\'{\i}n}, \citenamefont {Guratinder}, \citenamefont {Tsurkan}, \citenamefont {Prodan}, \citenamefont {Ressouche},\ and\ \citenamefont {Zaharko}}]{Rosales2022}%
  \BibitemOpen
  \bibfield  {author} {\bibinfo {author} {\bibfnamefont {H.~D.}\ \bibnamefont {Rosales}}, \bibinfo {author} {\bibfnamefont {F.~A.~G.}\ \bibnamefont {Albarrac\'{\i}n}}, \bibinfo {author} {\bibfnamefont {K.}~\bibnamefont {Guratinder}}, \bibinfo {author} {\bibfnamefont {V.}~\bibnamefont {Tsurkan}}, \bibinfo {author} {\bibfnamefont {L.}~\bibnamefont {Prodan}}, \bibinfo {author} {\bibfnamefont {E.}~\bibnamefont {Ressouche}}, \ and\ \bibinfo {author} {\bibfnamefont {O.}~\bibnamefont {Zaharko}},\ }\href {\doibase 10.1103/PhysRevB.105.224402} {\bibfield  {journal} {\bibinfo  {journal} {Physical Review B}\ }\textbf {\bibinfo {volume} {105}},\ \bibinfo {pages} {224402} (\bibinfo {year} {2022})}
  \BibitemShut {NoStop}%
\bibitem [{\citenamefont {Amoroso}\ \emph {et~al.}(2020)\citenamefont {Amoroso}, \citenamefont {Barone},\ and\ \citenamefont {Picozzi}}]{Amoroso2020}%
  \BibitemOpen
  \bibfield  {author} {\bibinfo {author} {\bibfnamefont {D.}~\bibnamefont {Amoroso}}, \bibinfo {author} {\bibfnamefont {P.}~\bibnamefont {Barone}}, \ and\ \bibinfo {author} {\bibfnamefont {S.}~\bibnamefont {Picozzi}},\ }\href {\doibase 10.1038/s41467-020-19535-w} {\bibfield  {journal} {\bibinfo  {journal} {Nature Communications}\ }\textbf {\bibinfo {volume} {11}} (\bibinfo {year} {2020}),\ 10.1038/s41467-020-19535-w}
  \BibitemShut {NoStop}%
\bibitem [{\citenamefont {Wang}\ \emph {et~al.}(2020)\citenamefont {Wang}, \citenamefont {Su}, \citenamefont {Lin},\ and\ \citenamefont {Batista}}]{Wang2020}%
  \BibitemOpen
  \bibfield  {author} {\bibinfo {author} {\bibfnamefont {Z.}~\bibnamefont {Wang}}, \bibinfo {author} {\bibfnamefont {Y.}~\bibnamefont {Su}}, \bibinfo {author} {\bibfnamefont {S.-Z.}\ \bibnamefont {Lin}}, \ and\ \bibinfo {author} {\bibfnamefont {C.~D.}\ \bibnamefont {Batista}},\ }\href {\doibase 10.1103/PhysRevLett.124.207201} {\bibfield  {journal} {\bibinfo  {journal} {Phys. Rev. Lett.}\ }\textbf {\bibinfo {volume} {124}},\ \bibinfo {pages} {207201} (\bibinfo {year} {2020})}
  \BibitemShut {NoStop}%
\bibitem [{\citenamefont {Utesov}(2021)}]{Utesov2021}%
  \BibitemOpen
  \bibfield  {author} {\bibinfo {author} {\bibfnamefont {O.~I.}\ \bibnamefont {Utesov}},\ }\href {https://doi.org/10.1103/PhysRevB.103.064414} {\bibfield  {journal} {\bibinfo  {journal} {Physical Review B}\ }\textbf {\bibinfo {volume} {103}},\ \bibinfo {pages} {064414} (\bibinfo {year} {2021})}
  \BibitemShut {NoStop}%
\bibitem [{\citenamefont {Berger}(1996)}]{Berger1996}%
  \BibitemOpen
  \bibfield  {author} {\bibinfo {author} {\bibfnamefont {L.}~\bibnamefont {Berger}},\ }\href {\doibase 10.1103/PhysRevB.54.9353} {\bibfield  {journal} {\bibinfo  {journal} {Phys. Rev. B}\ }\textbf {\bibinfo {volume} {54}},\ \bibinfo {pages} {9353} (\bibinfo {year} {1996})}
  \BibitemShut {NoStop}%
\bibitem [{\citenamefont {Slonczewski}(1996)}]{Slonczewski1996}%
  \BibitemOpen
  \bibfield  {author} {\bibinfo {author} {\bibfnamefont {J.}~\bibnamefont {Slonczewski}},\ }\href {\doibase https://doi.org/10.1016/0304-8853(96)00062-5} {\bibfield  {journal} {\bibinfo  {journal} {Journal of Magnetism and Magnetic Materials}\ }\textbf {\bibinfo {volume} {159}},\ \bibinfo {pages} {L1} (\bibinfo {year} {1996})}
  \BibitemShut {NoStop}%
\bibitem [{\citenamefont {Jonietz}\ \emph {et~al.}(2010)\citenamefont {Jonietz}, \citenamefont {MÃ¼hlbauer}, \citenamefont {Pfleiderer}, \citenamefont {Neubauer}, \citenamefont {MÃ¼nzer}, \citenamefont {Bauer}, \citenamefont {Adams}, \citenamefont {Georgii}, \citenamefont {BÃ¶ni}, \citenamefont {Duine}, \citenamefont {Everschor}, \citenamefont {Garst},\ and\ \citenamefont {Rosch}}]{Jonietz2010}%
  \BibitemOpen
  \bibfield  {author} {\bibinfo {author} {\bibfnamefont {F.}~\bibnamefont {Jonietz}}, \bibinfo {author} {\bibfnamefont {S.}~\bibnamefont {MÃ¼hlbauer}}, \bibinfo {author} {\bibfnamefont {C.}~\bibnamefont {Pfleiderer}}, \bibinfo {author} {\bibfnamefont {A.}~\bibnamefont {Neubauer}}, \bibinfo {author} {\bibfnamefont {W.}~\bibnamefont {MÃ¼nzer}}, \bibinfo {author} {\bibfnamefont {A.}~\bibnamefont {Bauer}}, \bibinfo {author} {\bibfnamefont {T.}~\bibnamefont {Adams}}, \bibinfo {author} {\bibfnamefont {R.}~\bibnamefont {Georgii}}, \bibinfo {author} {\bibfnamefont {P.}~\bibnamefont {BÃ¶ni}}, \bibinfo {author} {\bibfnamefont {R.~A.}\ \bibnamefont {Duine}}, \bibinfo {author} {\bibfnamefont {K.}~\bibnamefont {Everschor}}, \bibinfo {author} {\bibfnamefont {M.}~\bibnamefont {Garst}}, \ and\ \bibinfo {author} {\bibfnamefont {A.}~\bibnamefont {Rosch}},\ }\href {\doibase 10.1126/science.1195709} {\bibfield  {journal} {\bibinfo  {journal} {Science}\ }\textbf {\bibinfo {volume} {330}},\ \bibinfo {pages} {1648} (\bibinfo {year}
  {2010})}
  \BibitemShut {NoStop}%
\bibitem [{\citenamefont {Schulz}\ \emph {et~al.}(2012)\citenamefont {Schulz}, \citenamefont {Ritz}, \citenamefont {Bauer}, \citenamefont {Halder}, \citenamefont {Wagner}, \citenamefont {Franz}, \citenamefont {Pfleiderer}, \citenamefont {Everschor}, \citenamefont {Garst},\ and\ \citenamefont {Rosch}}]{Schulz2012}%
  \BibitemOpen
  \bibfield  {author} {\bibinfo {author} {\bibfnamefont {T.}~\bibnamefont {Schulz}}, \bibinfo {author} {\bibfnamefont {R.}~\bibnamefont {Ritz}}, \bibinfo {author} {\bibfnamefont {A.}~\bibnamefont {Bauer}}, \bibinfo {author} {\bibfnamefont {M.}~\bibnamefont {Halder}}, \bibinfo {author} {\bibfnamefont {M.~S.}\ \bibnamefont {Wagner}}, \bibinfo {author} {\bibfnamefont {C.}~\bibnamefont {Franz}}, \bibinfo {author} {\bibfnamefont {C.}~\bibnamefont {Pfleiderer}}, \bibinfo {author} {\bibfnamefont {K.}~\bibnamefont {Everschor}}, \bibinfo {author} {\bibfnamefont {M.}~\bibnamefont {Garst}}, \ and\ \bibinfo {author} {\bibfnamefont {A.}~\bibnamefont {Rosch}},\ }\href {https://api.semanticscholar.org/CorpusID:119119249} {\bibfield  {journal} {\bibinfo  {journal} {Nature Physics}\ }\textbf {\bibinfo {volume} {8}},\ \bibinfo {pages} {301 } (\bibinfo {year} {2012})}
  \BibitemShut {NoStop}%
\bibitem [{\citenamefont {Everschor}\ \emph {et~al.}(2011)\citenamefont {Everschor}, \citenamefont {Garst}, \citenamefont {Duine},\ and\ \citenamefont {Rosch}}]{Everschor2011}%
  \BibitemOpen
  \bibfield  {author} {\bibinfo {author} {\bibfnamefont {K.}~\bibnamefont {Everschor}}, \bibinfo {author} {\bibfnamefont {M.}~\bibnamefont {Garst}}, \bibinfo {author} {\bibfnamefont {R.~A.}\ \bibnamefont {Duine}}, \ and\ \bibinfo {author} {\bibfnamefont {A.}~\bibnamefont {Rosch}},\ }\href {\doibase 10.1103/PhysRevB.84.064401} {\bibfield  {journal} {\bibinfo  {journal} {Phys. Rev. B}\ }\textbf {\bibinfo {volume} {84}},\ \bibinfo {pages} {064401} (\bibinfo {year} {2011})}
  \BibitemShut {NoStop}%
\bibitem [{\citenamefont {Zang}\ \emph {et~al.}(2011)\citenamefont {Zang}, \citenamefont {Mostovoy}, \citenamefont {Han},\ and\ \citenamefont {Nagaosa}}]{Zang2011}%
  \BibitemOpen
  \bibfield  {author} {\bibinfo {author} {\bibfnamefont {J.}~\bibnamefont {Zang}}, \bibinfo {author} {\bibfnamefont {M.}~\bibnamefont {Mostovoy}}, \bibinfo {author} {\bibfnamefont {J.~H.}\ \bibnamefont {Han}}, \ and\ \bibinfo {author} {\bibfnamefont {N.}~\bibnamefont {Nagaosa}},\ }\href {\doibase 10.1103/PhysRevLett.107.136804} {\bibfield  {journal} {\bibinfo  {journal} {Phys. Rev. Lett.}\ }\textbf {\bibinfo {volume} {107}},\ \bibinfo {pages} {136804} (\bibinfo {year} {2011})}
  \BibitemShut {NoStop}%
\bibitem [{\citenamefont {Kimura}\ \emph {et~al.}(2003)\citenamefont {Kimura}, \citenamefont {Kimura}, \citenamefont {Goto}, \citenamefont {Shintani}, \citenamefont {Ishizaka}, \citenamefont {hisa Arima},\ and\ \citenamefont {Tokura}}]{Kimura2003}%
  \BibitemOpen
  \bibfield  {author} {\bibinfo {author} {\bibfnamefont {T.}~\bibnamefont {Kimura}}, \bibinfo {author} {\bibfnamefont {T.}~\bibnamefont {Kimura}}, \bibinfo {author} {\bibfnamefont {T.}~\bibnamefont {Goto}}, \bibinfo {author} {\bibfnamefont {H.}~\bibnamefont {Shintani}}, \bibinfo {author} {\bibfnamefont {K.}~\bibnamefont {Ishizaka}}, \bibinfo {author} {\bibfnamefont {T.}~\bibnamefont {hisa Arima}}, \ and\ \bibinfo {author} {\bibfnamefont {Y.}~\bibnamefont {Tokura}},\ }\href {https://api.semanticscholar.org/CorpusID:205209892} {\bibfield  {journal} {\bibinfo  {journal} {Nature}\ }\textbf {\bibinfo {volume} {426}},\ \bibinfo {pages} {55} (\bibinfo {year} {2003})}
  \BibitemShut {NoStop}%
\bibitem [{\citenamefont {Fiebig}(2005)}]{Fiebig2005}%
  \BibitemOpen
  \bibfield  {author} {\bibinfo {author} {\bibfnamefont {M.}~\bibnamefont {Fiebig}},\ }\href {\doibase 10.1088/0022-3727/38/8/R01} {\bibfield  {journal} {\bibinfo  {journal} {Journal of Physics D: Applied Physics}\ }\textbf {\bibinfo {volume} {38}},\ \bibinfo {pages} {R123} (\bibinfo {year} {2005})}
  \BibitemShut {NoStop}%
\bibitem [{\citenamefont {Tokura}(2006)}]{Tokura2006}%
  \BibitemOpen
  \bibfield  {author} {\bibinfo {author} {\bibfnamefont {Y.}~\bibnamefont {Tokura}},\ }\href {\doibase 10.1126/science.1125227} {\bibfield  {journal} {\bibinfo  {journal} {Science}\ }\textbf {\bibinfo {volume} {312}},\ \bibinfo {pages} {1481} (\bibinfo {year} {2006})}
  \BibitemShut {NoStop}%
\bibitem [{\citenamefont {Kimura}(2007)}]{Kimura2007}%
  \BibitemOpen
  \bibfield  {author} {\bibinfo {author} {\bibfnamefont {T.}~\bibnamefont {Kimura}},\ }\href {https://api.semanticscholar.org/CorpusID:137564504} {\bibfield  {journal} {\bibinfo  {journal} {Annual Review of Materials Research}\ }\textbf {\bibinfo {volume} {37}},\ \bibinfo {pages} {387} (\bibinfo {year} {2007})}
  \BibitemShut {NoStop}%
\bibitem [{\citenamefont {Cheong}\ and\ \citenamefont {Mostovoy}(2007)}]{Cheong2007}%
  \BibitemOpen
  \bibfield  {author} {\bibinfo {author} {\bibfnamefont {S.-W.}\ \bibnamefont {Cheong}}\ and\ \bibinfo {author} {\bibfnamefont {M.}~\bibnamefont {Mostovoy}},\ }\href {\doibase 10.1038/nmat1804} {\bibfield  {journal} {\bibinfo  {journal} {Nature Materials}\ }\textbf {\bibinfo {volume} {6}},\ \bibinfo {pages} {13} (\bibinfo {year} {2007})}
  \BibitemShut {NoStop}%
\bibitem [{\citenamefont {Tokura}\ and\ \citenamefont {Seki}(2010)}]{Tokura2010}%
  \BibitemOpen
  \bibfield  {author} {\bibinfo {author} {\bibfnamefont {Y.}~\bibnamefont {Tokura}}\ and\ \bibinfo {author} {\bibfnamefont {S.}~\bibnamefont {Seki}},\ }\href {https://api.semanticscholar.org/CorpusID:12517975} {\bibfield  {journal} {\bibinfo  {journal} {Advanced Materials}\ }\textbf {\bibinfo {volume} {22}} (\bibinfo {year} {2010})}
  \BibitemShut {NoStop}%
\bibitem [{\citenamefont {Khomskii}(2009)}]{Khomskii2009}%
  \BibitemOpen
  \bibfield  {author} {\bibinfo {author} {\bibfnamefont {D.}~\bibnamefont {Khomskii}},\ }\href@noop {} {\bibfield  {journal} {\bibinfo  {journal} {Physics}\ }\textbf {\bibinfo {volume} {2}} (\bibinfo {year} {2009})}
  \BibitemShut {NoStop}%
  \bibitem [{\citenamefont {Ruff}\ \emph {et~al.}(2015)\citenamefont {Ruff}, \citenamefont {Widmann}, \citenamefont {Lunkenheimer}, \citenamefont {Tsurkan}, \citenamefont {Bordács}, \citenamefont {Kézsmárki},\ and\ \citenamefont {Loidl}}]{Ruff2015}%
  \BibitemOpen
  \bibfield  {author} {\bibinfo {author} {\bibfnamefont {E.}~\bibnamefont {Ruff}}, \bibinfo {author} {\bibfnamefont {S.}~\bibnamefont {Widmann}}, \bibinfo {author} {\bibfnamefont {P.}~\bibnamefont {Lunkenheimer}}, \bibinfo {author} {\bibfnamefont {V.}~\bibnamefont {Tsurkan}}, \bibinfo {author} {\bibfnamefont {S.}~\bibnamefont {Bordács}}, \bibinfo {author} {\bibfnamefont {I.}~\bibnamefont {Kézsmárki}}, \ and\ \bibinfo {author} {\bibfnamefont {A.}~\bibnamefont {Loidl}},\ }\href {\doibase 10.1126/sciadv.1500916} {\bibfield  {journal} {\bibinfo  {journal} {Science Advances}\ }\textbf {\bibinfo {volume} {1}},\ \bibinfo {pages} {e1500916} (\bibinfo {year} {2015})}
  \BibitemShut {NoStop}%
\bibitem [{\citenamefont {Jia}\ \emph {et~al.}(2006)\citenamefont {Jia}, \citenamefont {Onoda}, \citenamefont {Nagaosa},\ and\ \citenamefont {Han}}]{Jia2006}%
  \BibitemOpen
  \bibfield  {author} {\bibinfo {author} {\bibfnamefont {C.}~\bibnamefont {Jia}}, \bibinfo {author} {\bibfnamefont {S.}~\bibnamefont {Onoda}}, \bibinfo {author} {\bibfnamefont {N.}~\bibnamefont {Nagaosa}}, \ and\ \bibinfo {author} {\bibfnamefont {J.~H.}\ \bibnamefont {Han}},\ }\href {\doibase 10.1103/PhysRevB.74.224444} {\bibfield  {journal} {\bibinfo  {journal} {Phys. Rev. B}\ }\textbf {\bibinfo {volume} {74}},\ \bibinfo {pages} {224444} (\bibinfo {year} {2006})}
  \BibitemShut {NoStop}%
\bibitem [{\citenamefont {Jia}\ \emph {et~al.}(2007)\citenamefont {Jia}, \citenamefont {Onoda}, \citenamefont {Nagaosa},\ and\ \citenamefont {Han}}]{Jia2007}%
  \BibitemOpen
  \bibfield  {author} {\bibinfo {author} {\bibfnamefont {C.}~\bibnamefont {Jia}}, \bibinfo {author} {\bibfnamefont {S.}~\bibnamefont {Onoda}}, \bibinfo {author} {\bibfnamefont {N.}~\bibnamefont {Nagaosa}}, \ and\ \bibinfo {author} {\bibfnamefont {J.~H.}\ \bibnamefont {Han}},\ }\href {\doibase 10.1103/PhysRevB.76.144424} {\bibfield  {journal} {\bibinfo  {journal} {Phys. Rev. B}\ }\textbf {\bibinfo {volume} {76}},\ \bibinfo {pages} {144424} (\bibinfo {year} {2007})}
  \BibitemShut {NoStop}%
\bibitem [{\citenamefont {Murakawa}\ \emph {et~al.}(2010)\citenamefont {Murakawa}, \citenamefont {Onose}, \citenamefont {Miyahara}, \citenamefont {Furukawa},\ and\ \citenamefont {Tokura}}]{Murakawa2010}%
  \BibitemOpen
  \bibfield  {author} {\bibinfo {author} {\bibfnamefont {H.}~\bibnamefont {Murakawa}}, \bibinfo {author} {\bibfnamefont {Y.}~\bibnamefont {Onose}}, \bibinfo {author} {\bibfnamefont {S.}~\bibnamefont {Miyahara}}, \bibinfo {author} {\bibfnamefont {N.}~\bibnamefont {Furukawa}}, \ and\ \bibinfo {author} {\bibfnamefont {Y.}~\bibnamefont {Tokura}},\ }\href {\doibase 10.1103/PhysRevLett.105.137202} {\bibfield  {journal} {\bibinfo  {journal} {Phys. Rev. Lett.}\ }\textbf {\bibinfo {volume} {105}},\ \bibinfo {pages} {137202} (\bibinfo {year} {2010})}
  \BibitemShut {NoStop}%
\bibitem [{\citenamefont {Katsura}\ \emph {et~al.}(2005)\citenamefont {Katsura}, \citenamefont {Nagaosa},\ and\ \citenamefont {Balatsky}}]{Katsura2005}%
  \BibitemOpen
  \bibfield  {author} {\bibinfo {author} {\bibfnamefont {H.}~\bibnamefont {Katsura}}, \bibinfo {author} {\bibfnamefont {N.}~\bibnamefont {Nagaosa}}, \ and\ \bibinfo {author} {\bibfnamefont {A.~V.}\ \bibnamefont {Balatsky}},\ }\href {\doibase 10.1103/PhysRevLett.95.057205} {\bibfield  {journal} {\bibinfo  {journal} {Phys. Rev. Lett.}\ }\textbf {\bibinfo {volume} {95}},\ \bibinfo {pages} {057205} (\bibinfo {year} {2005})}
  \BibitemShut {NoStop}%
\bibitem [{\citenamefont {Mostovoy}(2006)}]{Mostovoy2006}%
  \BibitemOpen
  \bibfield  {author} {\bibinfo {author} {\bibfnamefont {M.}~\bibnamefont {Mostovoy}},\ }\href {\doibase 10.1103/PhysRevLett.96.067601} {\bibfield  {journal} {\bibinfo  {journal} {Phys. Rev. Lett.}\ }\textbf {\bibinfo {volume} {96}},\ \bibinfo {pages} {067601} (\bibinfo {year} {2006})}
  \BibitemShut {NoStop}%
\bibitem [{\citenamefont {Sergienko}\ and\ \citenamefont {Dagotto}(2006)}]{Sergienko2006}%
  \BibitemOpen
  \bibfield  {author} {\bibinfo {author} {\bibfnamefont {I.~A.}\ \bibnamefont {Sergienko}}\ and\ \bibinfo {author} {\bibfnamefont {E.}~\bibnamefont {Dagotto}},\ }\href {\doibase 10.1103/PhysRevB.73.094434} {\bibfield  {journal} {\bibinfo  {journal} {Phys. Rev. B}\ }\textbf {\bibinfo {volume} {73}},\ \bibinfo {pages} {094434} (\bibinfo {year} {2006})}
  \BibitemShut {NoStop}%
\bibitem [{\citenamefont {Arima}\ \emph {et~al.}(2006)\citenamefont {Arima}, \citenamefont {Tokunaga}, \citenamefont {Goto}, \citenamefont {Kimura}, \citenamefont {Noda},\ and\ \citenamefont {Tokura}}]{Arima2006}%
  \BibitemOpen
  \bibfield  {author} {\bibinfo {author} {\bibfnamefont {T.}~\bibnamefont {Arima}}, \bibinfo {author} {\bibfnamefont {A.}~\bibnamefont {Tokunaga}}, \bibinfo {author} {\bibfnamefont {T.}~\bibnamefont {Goto}}, \bibinfo {author} {\bibfnamefont {H.}~\bibnamefont {Kimura}}, \bibinfo {author} {\bibfnamefont {Y.}~\bibnamefont {Noda}}, \ and\ \bibinfo {author} {\bibfnamefont {Y.}~\bibnamefont {Tokura}},\ }\href {\doibase 10.1103/PhysRevLett.96.097202} {\bibfield  {journal} {\bibinfo  {journal} {Phys. Rev. Lett.}\ }\textbf {\bibinfo {volume} {96}},\ \bibinfo {pages} {097202} (\bibinfo {year} {2006})}  
  \BibitemShut {NoStop}%
\bibitem [{\citenamefont {Choi}\ \emph {et~al.}(2008)\citenamefont {Choi}, \citenamefont {Yi}, \citenamefont {Lee}, \citenamefont {Huang}, \citenamefont {Kiryukhin},\ and\ \citenamefont {Cheong}}]{Choi2008}%
  \BibitemOpen
  \bibfield  {author} {\bibinfo {author} {\bibfnamefont {Y.~J.}\ \bibnamefont {Choi}}, \bibinfo {author} {\bibfnamefont {H.~T.}\ \bibnamefont {Yi}}, \bibinfo {author} {\bibfnamefont {S.}~\bibnamefont {Lee}}, \bibinfo {author} {\bibfnamefont {Q.}~\bibnamefont {Huang}}, \bibinfo {author} {\bibfnamefont {V.}~\bibnamefont {Kiryukhin}}, \ and\ \bibinfo {author} {\bibfnamefont {S.-W.}\ \bibnamefont {Cheong}},\ }\href {\doibase 10.1103/PhysRevLett.100.047601} {\bibfield  {journal} {\bibinfo  {journal} {Phys. Rev. Lett.}\ }\textbf {\bibinfo {volume} {100}},\ \bibinfo {pages} {047601} (\bibinfo {year} {2008})}
  \BibitemShut {NoStop}%
\bibitem [{\citenamefont {Tokunaga}\ \emph {et~al.}(2008)\citenamefont {Tokunaga}, \citenamefont {Iguchi}, \citenamefont {Arima},\ and\ \citenamefont {Tokura}}]{Tokunaga2008}%
  \BibitemOpen
  \bibfield  {author} {\bibinfo {author} {\bibfnamefont {Y.}~\bibnamefont {Tokunaga}}, \bibinfo {author} {\bibfnamefont {S.}~\bibnamefont {Iguchi}}, \bibinfo {author} {\bibfnamefont {T.}~\bibnamefont {Arima}}, \ and\ \bibinfo {author} {\bibfnamefont {Y.}~\bibnamefont {Tokura}},\ }\href {\doibase 10.1103/PhysRevLett.101.097205} {\bibfield  {journal} {\bibinfo  {journal} {Phys. Rev. Lett.}\ }\textbf {\bibinfo {volume} {101}},\ \bibinfo {pages} {097205} (\bibinfo {year} {2008})}
  \BibitemShut {NoStop}%
\bibitem [{\citenamefont {Ishiwata}\ \emph {et~al.}(2010)\citenamefont {Ishiwata}, \citenamefont {Kaneko}, \citenamefont {Tokunaga}, \citenamefont {Taguchi}, \citenamefont {Arima},\ and\ \citenamefont {Tokura}}]{Ishiwata2010}%
  \BibitemOpen
  \bibfield  {author} {\bibinfo {author} {\bibfnamefont {S.}~\bibnamefont {Ishiwata}}, \bibinfo {author} {\bibfnamefont {Y.}~\bibnamefont {Kaneko}}, \bibinfo {author} {\bibfnamefont {Y.}~\bibnamefont {Tokunaga}}, \bibinfo {author} {\bibfnamefont {Y.}~\bibnamefont {Taguchi}}, \bibinfo {author} {\bibfnamefont {T.-h.}\ \bibnamefont {Arima}}, \ and\ \bibinfo {author} {\bibfnamefont {Y.}~\bibnamefont {Tokura}},\ }\href {\doibase 10.1103/PhysRevB.81.100411} {\bibfield  {journal} {\bibinfo  {journal} {Phys. Rev. B}\ }\textbf {\bibinfo {volume} {81}},\ \bibinfo {pages} {100411} (\bibinfo {year} {2010})}   
  \BibitemShut {NoStop}%
\bibitem [{\citenamefont {Cabra}\ and\ \citenamefont {Rossini}(2025)}]{Cabra2025}%
  \BibitemOpen
  \bibfield  {author} {\bibinfo {author} {\bibfnamefont {D.~C.}\ \bibnamefont {Cabra}}\ and\ \bibinfo {author} {\bibfnamefont {G.~L.}\ \bibnamefont {Rossini}},\ }\href {\doibase 10.1063/5.0273516} {\bibfield  {journal} {\bibinfo  {journal} {Journal of Applied Physics}\ }\textbf {\bibinfo {volume} {138}} (\bibinfo {year} {2025}),\ 10.1063/5.0273516}
  \BibitemShut {NoStop}%
\bibitem [{\citenamefont {Arima}(2007)}]{Arima2007}%
  \BibitemOpen
  \bibfield  {author} {\bibinfo {author} {\bibfnamefont {T.-h.}\ \bibnamefont {Arima}},\ }\href@noop {} {\bibfield  {journal} {\bibinfo  {journal} {J PHYS SOC JPN}\ }\textbf {\bibinfo {volume} {76}} (\bibinfo {year} {2007})}
  \BibitemShut {NoStop}%
\bibitem [{\citenamefont {Murakawa}\ \emph {et~al.}(2012)\citenamefont {Murakawa}, \citenamefont {Onose}, \citenamefont {Miyahara}, \citenamefont {Furukawa},\ and\ \citenamefont {Tokura}}]{Murakawa2012}%
  \BibitemOpen
  \bibfield  {author} {\bibinfo {author} {\bibfnamefont {H.}~\bibnamefont {Murakawa}}, \bibinfo {author} {\bibfnamefont {Y.}~\bibnamefont {Onose}}, \bibinfo {author} {\bibfnamefont {S.}~\bibnamefont {Miyahara}}, \bibinfo {author} {\bibfnamefont {N.}~\bibnamefont {Furukawa}}, \ and\ \bibinfo {author} {\bibfnamefont {Y.}~\bibnamefont {Tokura}},\ }\href {\doibase 10.1103/PhysRevB.85.174106} {\bibfield  {journal} {\bibinfo  {journal} {Phys. Rev. B}\ }\textbf {\bibinfo {volume} {85}},\ \bibinfo {pages} {174106} (\bibinfo {year} {2012})}
  \BibitemShut {NoStop}%
\bibitem [{\citenamefont {Seki}\ \emph {et~al.}(2012{\natexlab{a}})\citenamefont {Seki}, \citenamefont {yu}, \citenamefont {Ishiwata},\ and\ \citenamefont {Tokura}}]{Seki2012a}%
  \BibitemOpen
  \bibfield  {author} {\bibinfo {author} {\bibfnamefont {S.}~\bibnamefont {Seki}}, \bibinfo {author} {\bibfnamefont {X.}~\bibnamefont {yu}}, \bibinfo {author} {\bibfnamefont {S.}~\bibnamefont {Ishiwata}}, \ and\ \bibinfo {author} {\bibfnamefont {Y.}~\bibnamefont {Tokura}},\ }\href {\doibase 10.1126/science.1214143} {\bibfield  {journal} {\bibinfo  {journal} {Science (New York, N.Y.)}\ }\textbf {\bibinfo {volume} {336}},\ \bibinfo {pages} {198} (\bibinfo {year} {2012}{\natexlab{a}})}
  \BibitemShut {NoStop}%
\bibitem [{\citenamefont {Liu}\ \emph {et~al.}(2013{\natexlab{a}})\citenamefont {Liu}, \citenamefont {Li},\ and\ \citenamefont {Han}}]{Liu2013a}%
  \BibitemOpen
  \bibfield  {author} {\bibinfo {author} {\bibfnamefont {Y.-H.}\ \bibnamefont {Liu}}, \bibinfo {author} {\bibfnamefont {Y.-Q.}\ \bibnamefont {Li}}, \ and\ \bibinfo {author} {\bibfnamefont {J.~H.}\ \bibnamefont {Han}},\ }\href {\doibase 10.1103/PhysRevB.87.100402} {\bibfield  {journal} {\bibinfo  {journal} {Phys. Rev. B}\ }\textbf {\bibinfo {volume} {87}},\ \bibinfo {pages} {100402} (\bibinfo {year} {2013}{\natexlab{a}})}
  \BibitemShut {NoStop}%
\bibitem [{\citenamefont {Liu}\ \emph {et~al.}(2013{\natexlab{b}})\citenamefont {Liu}, \citenamefont {Hoon}, \citenamefont {Alahgholipour~Omrani}, \citenamefont {Ronnow},\ and\ \citenamefont {Li}}]{Liu2013b}%
  \BibitemOpen
  \bibfield  {author} {\bibinfo {author} {\bibfnamefont {Y.-H.}\ \bibnamefont {Liu}}, \bibinfo {author} {\bibfnamefont {J.}~\bibnamefont {Hoon}}, \bibinfo {author} {\bibfnamefont {A.}~\bibnamefont {Alahgholipour~Omrani}}, \bibinfo {author} {\bibfnamefont {H.}~\bibnamefont {Ronnow}}, \ and\ \bibinfo {author} {\bibfnamefont {Y.-Q.}\ \bibnamefont {Li}},\ }\href@noop {} {\bibfield  {journal} {\bibinfo  {journal} {arxiv}\ }\textbf {\bibinfo {volume} {70}} (\bibinfo {year} {2013}{\natexlab{b}})}
  \BibitemShut {NoStop}%
\bibitem [{\citenamefont {Kimura}\ \emph {et~al.}(2006)\citenamefont {Kimura}, \citenamefont {Lashley},\ and\ \citenamefont {Ramirez}}]{Kimura2006}%
  \BibitemOpen
  \bibfield  {author} {\bibinfo {author} {\bibfnamefont {T.}~\bibnamefont {Kimura}}, \bibinfo {author} {\bibfnamefont {J.~C.}\ \bibnamefont {Lashley}}, \ and\ \bibinfo {author} {\bibfnamefont {A.~P.}\ \bibnamefont {Ramirez}},\ }\href {\doibase 10.1103/PhysRevB.73.220401} {\bibfield  {journal} {\bibinfo  {journal} {Phys. Rev. B}\ }\textbf {\bibinfo {volume} {73}},\ \bibinfo {pages} {220401} (\bibinfo {year} {2006})}
  \BibitemShut {NoStop}%
\bibitem [{\citenamefont {Seki}\ \emph {et~al.}(2008)\citenamefont {Seki}, \citenamefont {Onose},\ and\ \citenamefont {Tokura}}]{Seki2008}%
  \BibitemOpen
  \bibfield  {author} {\bibinfo {author} {\bibfnamefont {S.}~\bibnamefont {Seki}}, \bibinfo {author} {\bibfnamefont {Y.}~\bibnamefont {Onose}}, \ and\ \bibinfo {author} {\bibfnamefont {Y.}~\bibnamefont {Tokura}},\ }\href {\doibase 10.1103/PhysRevLett.101.067204} {\bibfield  {journal} {\bibinfo  {journal} {Phys. Rev. Lett.}\ }\textbf {\bibinfo {volume} {101}},\ \bibinfo {pages} {067204} (\bibinfo {year} {2008})}
  \BibitemShut {NoStop}%
\bibitem [{\citenamefont {Belesi}\ \emph {et~al.}(2012)\citenamefont {Belesi}, \citenamefont {Rousochatzakis}, \citenamefont {Abid}, \citenamefont {R\"o\ss{}ler}, \citenamefont {Berger},\ and\ \citenamefont {Ansermet}}]{Belesi2012}%
  \BibitemOpen
  \bibfield  {author} {\bibinfo {author} {\bibfnamefont {M.}~\bibnamefont {Belesi}}, \bibinfo {author} {\bibfnamefont {I.}~\bibnamefont {Rousochatzakis}}, \bibinfo {author} {\bibfnamefont {M.}~\bibnamefont {Abid}}, \bibinfo {author} {\bibfnamefont {U.~K.}\ \bibnamefont {R\"o\ss{}ler}}, \bibinfo {author} {\bibfnamefont {H.}~\bibnamefont {Berger}}, \ and\ \bibinfo {author} {\bibfnamefont {J.-P.}\ \bibnamefont {Ansermet}},\ }\href {\doibase 10.1103/PhysRevB.85.224413} {\bibfield  {journal} {\bibinfo  {journal} {Phys. Rev. B}\ }\textbf {\bibinfo {volume} {85}},\ \bibinfo {pages} {224413} (\bibinfo {year} {2012})}
  \BibitemShut {NoStop}%
\bibitem [{\citenamefont {Seki}\ \emph {et~al.}(2012{\natexlab{b}})\citenamefont {Seki}, \citenamefont {Ishiwata},\ and\ \citenamefont {Tokura}}]{Seki2012b}%
  \BibitemOpen
  \bibfield  {author} {\bibinfo {author} {\bibfnamefont {S.}~\bibnamefont {Seki}}, \bibinfo {author} {\bibfnamefont {S.}~\bibnamefont {Ishiwata}}, \ and\ \bibinfo {author} {\bibfnamefont {Y.}~\bibnamefont {Tokura}},\ }\href {\doibase 10.1103/PhysRevB.86.060403} {\bibfield  {journal} {\bibinfo  {journal} {Phys. Rev. B}\ }\textbf {\bibinfo {volume} {86}},\ \bibinfo {pages} {060403} (\bibinfo {year} {2012}{\natexlab{b}})}
  \BibitemShut {NoStop}%
\bibitem [{\citenamefont {White}\ \emph {et~al.}(2014)\citenamefont {White}, \citenamefont {Pr\ifmmode~\check{s}\else \v{s}\fi{}a}, \citenamefont {Huang}, \citenamefont {Omrani}, \citenamefont {\ifmmode \check{Z}\else \v{Z}\fi{}ivkovi\ifmmode~\acute{c}\else \'{c}\fi{}}, \citenamefont {Bartkowiak}, \citenamefont {Berger}, \citenamefont {Magrez}, \citenamefont {Gavilano}, \citenamefont {Nagy}, \citenamefont {Zang},\ and\ \citenamefont {R\o{}nnow}}]{White2012}%
  \BibitemOpen
  \bibfield  {author} {\bibinfo {author} {\bibfnamefont {J.~S.}\ \bibnamefont {White}}, \bibinfo {author} {\bibfnamefont {K.}~\bibnamefont {Pr\ifmmode~\check{s}\else \v{s}\fi{}a}}, \bibinfo {author} {\bibfnamefont {P.}~\bibnamefont {Huang}}, \bibinfo {author} {\bibfnamefont {A.~A.}\ \bibnamefont {Omrani}}, \bibinfo {author} {\bibfnamefont {I.}~\bibnamefont {\ifmmode \check{Z}\else \v{Z}\fi{}ivkovi\ifmmode~\acute{c}\else \'{c}\fi{}}}, \bibinfo {author} {\bibfnamefont {M.}~\bibnamefont {Bartkowiak}}, \bibinfo {author} {\bibfnamefont {H.}~\bibnamefont {Berger}}, \bibinfo {author} {\bibfnamefont {A.}~\bibnamefont {Magrez}}, \bibinfo {author} {\bibfnamefont {J.~L.}\ \bibnamefont {Gavilano}}, \bibinfo {author} {\bibfnamefont {G.}~\bibnamefont {Nagy}}, \bibinfo {author} {\bibfnamefont {J.}~\bibnamefont {Zang}}, \ and\ \bibinfo {author} {\bibfnamefont {H.~M.}\ \bibnamefont {R\o{}nnow}},\ }\href {\doibase 10.1103/PhysRevLett.113.107203} {\bibfield  {journal} {\bibinfo  {journal} {Phys. Rev. Lett.}\ }\textbf {\bibinfo
  {volume} {113}},\ \bibinfo {pages} {107203} (\bibinfo {year} {2014})}
  \BibitemShut {NoStop}%
\bibitem [{\citenamefont {White}\ \emph {et~al.}(2012)\citenamefont {White}, \citenamefont {LevatiÄ‡}, \citenamefont {Alahgholipour~Omrani}, \citenamefont {Egetemeyer}, \citenamefont {Prsa}, \citenamefont {Zivkovic}, \citenamefont {Gavilano}, \citenamefont {Kohlbrecher}, \citenamefont {Bartkowiak}, \citenamefont {Berger},\ and\ \citenamefont {Ronnow}}]{White2014}%
  \BibitemOpen
  \bibfield  {author} {\bibinfo {author} {\bibfnamefont {J.}~\bibnamefont {White}}, \bibinfo {author} {\bibfnamefont {I.}~\bibnamefont {LevatiÄ‡}}, \bibinfo {author} {\bibfnamefont {A.}~\bibnamefont {Alahgholipour~Omrani}}, \bibinfo {author} {\bibfnamefont {N.}~\bibnamefont {Egetemeyer}}, \bibinfo {author} {\bibfnamefont {K.}~\bibnamefont {Prsa}}, \bibinfo {author} {\bibfnamefont {I.}~\bibnamefont {Zivkovic}}, \bibinfo {author} {\bibfnamefont {J.}~\bibnamefont {Gavilano}}, \bibinfo {author} {\bibfnamefont {J.}~\bibnamefont {Kohlbrecher}}, \bibinfo {author} {\bibfnamefont {M.}~\bibnamefont {Bartkowiak}}, \bibinfo {author} {\bibfnamefont {H.}~\bibnamefont {Berger}}, \ and\ \bibinfo {author} {\bibfnamefont {H.}~\bibnamefont {Ronnow}},\ }\href@noop {} {\bibfield  {journal} {\bibinfo  {journal} {Journal of Physics Condensed Matter}\ }\textbf {\bibinfo {volume} {24}},\ \bibinfo {pages} {432201} (\bibinfo {year} {2012})}
  \BibitemShut {NoStop}%
\bibitem [{\citenamefont {Mochizuki}\ and\ \citenamefont {Watanabe}(2015)}]{Mochizuki2015a}%
  \BibitemOpen
  \bibfield  {author} {\bibinfo {author} {\bibfnamefont {M.}~\bibnamefont {Mochizuki}}\ and\ \bibinfo {author} {\bibfnamefont {Y.}~\bibnamefont {Watanabe}},\ }\href {\doibase 10.1063/1.4929727} {\bibfield  {journal} {\bibinfo  {journal} {Applied Physics Letters}\ }\textbf {\bibinfo {volume} {107}},\ \bibinfo {pages} {082409} (\bibinfo {year} {2015})}
  \BibitemShut {NoStop}%
\bibitem [{\citenamefont {Mochizuki}\ and\ \citenamefont {Seki}(2015)}]{Mochizuki2015b}%
  \BibitemOpen
  \bibfield  {author} {\bibinfo {author} {\bibfnamefont {M.}~\bibnamefont {Mochizuki}}\ and\ \bibinfo {author} {\bibfnamefont {S.}~\bibnamefont {Seki}},\ }\href {\doibase 10.1088/0953-8984/27/50/503001} {\bibfield  {journal} {\bibinfo  {journal} {Journal of Physics: Condensed Matter}\ }\textbf {\bibinfo {volume} {27}},\ \bibinfo {pages} {503001} (\bibinfo {year} {2015})}
  \BibitemShut {NoStop}%
\bibitem [{\citenamefont {Mochizuki}(2016)}]{Mochizuki2016}%
  \BibitemOpen
  \bibfield  {author} {\bibinfo {author} {\bibfnamefont {M.}~\bibnamefont {Mochizuki}},\ }\href@noop {} {\bibfield  {journal} {\bibinfo  {journal} {Advanced Electronic Materials}\ }\textbf {\bibinfo {volume} {2}},\ \bibinfo {pages} {1500180} (\bibinfo {year} {2016})}
  \BibitemShut {NoStop}%
\bibitem [{\citenamefont {Okamura}\ \emph {et~al.}(2016)\citenamefont {Okamura}, \citenamefont {Kagawa}, \citenamefont {Seki}, \citenamefont {Seki}, \citenamefont {Tokura},\ and\ \citenamefont {Tokura}}]{Okamura2016}%
  \BibitemOpen
  \bibfield  {author} {\bibinfo {author} {\bibfnamefont {Y.}~\bibnamefont {Okamura}}, \bibinfo {author} {\bibfnamefont {F.}~\bibnamefont {Kagawa}}, \bibinfo {author} {\bibfnamefont {S.}~\bibnamefont {Seki}}, \bibinfo {author} {\bibfnamefont {S.}~\bibnamefont {Seki}}, \bibinfo {author} {\bibfnamefont {Y.}~\bibnamefont {Tokura}}, \ and\ \bibinfo {author} {\bibfnamefont {Y.}~\bibnamefont {Tokura}},\ }\href {https://api.semanticscholar.org/CorpusID:18382591} {\bibfield  {journal} {\bibinfo  {journal} {Nature Communications}\ }\textbf {\bibinfo {volume} {7}} (\bibinfo {year} {2016})}
  \BibitemShut {NoStop}%
\bibitem [{\citenamefont {Kruchkov}\ \emph {et~al.}(2017)\citenamefont {Kruchkov}, \citenamefont {White}, \citenamefont {Bartowiak}, \citenamefont {Zivcovic}, \citenamefont {Magrez},\ and\ \citenamefont {Ronnow}}]{Kruchkov2017}%
  \BibitemOpen
  \bibfield  {author} {\bibinfo {author} {\bibfnamefont {A.}~\bibnamefont {Kruchkov}}, \bibinfo {author} {\bibfnamefont {J.}~\bibnamefont {White}}, \bibinfo {author} {\bibfnamefont {M.}~\bibnamefont {Bartowiak}}, \bibinfo {author} {\bibfnamefont {I.}~\bibnamefont {Zivcovic}}, \bibinfo {author} {\bibfnamefont {A.}~\bibnamefont {Magrez}}, \ and\ \bibinfo {author} {\bibfnamefont {H.}~\bibnamefont {Ronnow}},\ }\href@noop {} {\bibfield  {journal} {\bibinfo  {journal} {Scientific Reports}\ }\textbf {\bibinfo {volume} {8}} (\bibinfo {year} {2017})}
  \BibitemShut {NoStop}%
\bibitem [{\citenamefont {G\'omez~Albarrac\'{\i}n}\ and\ \citenamefont {Rosales}(2022)}]{Albarracin2022}%
  \BibitemOpen
  \bibfield  {author} {\bibinfo {author} {\bibfnamefont {F.~A.}\ \bibnamefont {G\'omez~Albarrac\'{\i}n}}\ and\ \bibinfo {author} {\bibfnamefont {H.~D.}\ \bibnamefont {Rosales}},\ }\href {\doibase 10.1103/PhysRevB.105.214423} {\bibfield  {journal} {\bibinfo  {journal} {Phys. Rev. B}\ }\textbf {\bibinfo {volume} {105}},\ \bibinfo {pages} {214423} (\bibinfo {year} {2022})}\BibitemShut {NoStop}%
\bibitem [{\citenamefont {Ezawa}(2011)}]{Ezawa2011}%
  \BibitemOpen
  \bibfield  {author} {\bibinfo {author} {\bibfnamefont {M.}~\bibnamefont {Ezawa}},\ }\href {\doibase 10.1103/PhysRevB.83.100408} {\bibfield  {journal} {\bibinfo  {journal} {Phys. Rev. B}\ }\textbf {\bibinfo {volume} {83}},\ \bibinfo {pages} {100408} (\bibinfo {year} {2011})}
  \BibitemShut {NoStop}%
\bibitem [{\citenamefont {K\'ezsm\'arki}\ \emph {et~al.}(2015)\citenamefont {K\'ezsm\'arki}, \citenamefont {Bord\'acs}, \citenamefont {Milde}, \citenamefont {Neuber}, \citenamefont {Eng}, \citenamefont {White}, \citenamefont {Ronnow}, \citenamefont {Dewhurst}, \citenamefont {Mochizuki}, \citenamefont {Yanai}, \citenamefont {Nakamura}, \citenamefont {Ehlers}, \citenamefont {Tsurkan},\ and\ \citenamefont {Loidl}}]{Kezsmarki2015}%
  \BibitemOpen
  \bibfield  {author} {\bibinfo {author} {\bibfnamefont {I.}~\bibnamefont {K\'ezsm\'arki}}, \bibinfo {author} {\bibfnamefont {S.}~\bibnamefont {Bordács}}, \bibinfo {author} {\bibfnamefont {P.}~\bibnamefont {Milde}}, \bibinfo {author} {\bibfnamefont {E.}~\bibnamefont {Neuber}}, \bibinfo {author} {\bibfnamefont {L.}~\bibnamefont {Eng}}, \bibinfo {author} {\bibfnamefont {J.}~\bibnamefont {White}}, \bibinfo {author} {\bibfnamefont {H.}~\bibnamefont {Ronnow}}, \bibinfo {author} {\bibfnamefont {C.}~\bibnamefont {Dewhurst}}, \bibinfo {author} {\bibfnamefont {M.}~\bibnamefont {Mochizuki}}, \bibinfo {author} {\bibfnamefont {K.}~\bibnamefont {Yanai}}, \bibinfo {author} {\bibfnamefont {H.}~\bibnamefont {Nakamura}}, \bibinfo {author} {\bibfnamefont {D.}~\bibnamefont {Ehlers}}, \bibinfo {author} {\bibfnamefont {V.}~\bibnamefont {Tsurkan}}, \ and\ \bibinfo {author} {\bibfnamefont {A.}~\bibnamefont {Loidl}},\ }\href@noop {} {\bibfield  {journal} {\bibinfo  {journal} {Nature materials}\ }\textbf {\bibinfo {volume} {14}}
  (\bibinfo {year} {2015})}
  \BibitemShut {NoStop}%
\bibitem [{\citenamefont {Bordács}\ \emph {et~al.}(2017)\citenamefont {Bordács}, \citenamefont {Butykai}, \citenamefont {Szigeti}, \citenamefont {White}, \citenamefont {Cubitt}, \citenamefont {Leonov}, \citenamefont {Widmann}, \citenamefont {Ehlers}, \citenamefont {Nidda}, \citenamefont {Tsurkan}, \citenamefont {Loidl},\ and\ \citenamefont {Kézsmárki}}]{Bordacs2017}%
  \BibitemOpen
  \bibfield  {author} {\bibinfo {author} {\bibfnamefont {S.}~\bibnamefont {Bordács}}, \bibinfo {author} {\bibfnamefont {Ã.}~\bibnamefont {Butykai}}, \bibinfo {author} {\bibfnamefont {B.}~\bibnamefont {Szigeti}}, \bibinfo {author} {\bibfnamefont {J.}~\bibnamefont {White}}, \bibinfo {author} {\bibfnamefont {R.}~\bibnamefont {Cubitt}}, \bibinfo {author} {\bibfnamefont {A.}~\bibnamefont {Leonov}}, \bibinfo {author} {\bibfnamefont {S.}~\bibnamefont {Widmann}}, \bibinfo {author} {\bibfnamefont {D.}~\bibnamefont {Ehlers}}, \bibinfo {author} {\bibfnamefont {H.-A.}\ \bibnamefont {Nidda}}, \bibinfo {author} {\bibfnamefont {V.}~\bibnamefont {Tsurkan}}, \bibinfo {author} {\bibfnamefont {A.}~\bibnamefont {Loidl}}, \ and\ \bibinfo {author} {\bibfnamefont {I.}~\bibnamefont {Kézsmárki}},\ }\href {\doibase 10.1038/s41598-017-07996-x} {\bibfield  {journal} {\bibinfo  {journal} {Scientific Reports}\ }\textbf {\bibinfo {volume} {7}} (\bibinfo {year} {2017})}
  \BibitemShut {NoStop}%
\end{thebibliography}

\end{document}